\documentclass{emulateapj}

\shorttitle{BOSS 1D}
\shortauthors{Bolton et al.}
 
\begin{document}

\title{Spectral Classification and Redshift Measurement for the SDSS-III Baryon Oscillation Spectroscopic Survey}

\author{\mbox{Adam S. Bolton\altaffilmark{1}}}
\author{\mbox{David J. Schlegel\altaffilmark{2}}}
\author{\mbox{\'{E}ric Aubourg\altaffilmark{3,4}}}
\author{\mbox{Stephen Bailey\altaffilmark{2}}}
\author{\mbox{Vaishali Bhardwaj\altaffilmark{5}}}
\author{\mbox{Joel R. Brownstein\altaffilmark{1}}}
\author{\mbox{Scott Burles\altaffilmark{6}}}
\author{\mbox{Yan-Mei Chen\altaffilmark{7}}}
\author{\mbox{Kyle Dawson\altaffilmark{1}}}
\author{\mbox{Daniel J. Eisenstein\altaffilmark{8}}}
\author{\mbox{James E. Gunn\altaffilmark{9}}}
\author{\mbox{G. R. Knapp\altaffilmark{9}}}
\author{\mbox{Craig P. Loomis\altaffilmark{9}}}
\author{\mbox{Robert H. Lupton\altaffilmark{9}}}
\author{\mbox{Claudia Maraston\altaffilmark{10}}}
\author{\mbox{Demitri Muna\altaffilmark{11}}}
\author{\mbox{Adam D. Myers\altaffilmark{12}}}
\author{\mbox{Matthew D. Olmstead\altaffilmark{1}}}
\author{\mbox{Nikhil Padmanabhan\altaffilmark{13}}}
\author{\mbox{Isabelle P\^{a}ris\altaffilmark{14,15}}}
\author{\mbox{Will J. Percival\altaffilmark{10}}}
\author{\mbox{Patrick Petitjean\altaffilmark{14}}}
\author{\mbox{Constance M. Rockosi\altaffilmark{16}}}
\author{\mbox{Nicholas P. Ross\altaffilmark{2}}}
\author{\mbox{Donald P. Schneider\altaffilmark{17,18}}}
\author{\mbox{Yiping Shu\altaffilmark{1}}}
\author{\mbox{Michael A. Strauss\altaffilmark{9}}}
\author{\mbox{Daniel Thomas\altaffilmark{10}}}
\author{\mbox{Christy A. Tremonti\altaffilmark{7}}}
\author{\mbox{David A. Wake\altaffilmark{19}}}
\author{\mbox{Benjamin A. Weaver\altaffilmark{11}}}
\author{\mbox{W. Michael Wood-Vasey\altaffilmark{20}}}

\altaffiltext{1}{Department of Physics and Astronomy, University of Utah,
115 South 1400 East, Salt Lake City, UT 84112, USA ({\tt bolton@astro.utah.edu})}
\altaffiltext{2}{Lawrence Berkeley National Laboratory, 1 Cyclotron Rd., Berkeley, CA 94720, USA}
\altaffiltext{3}{Astroparticule et Cosmologie (APC), Universit\'{e} Paris-Diderot, 75205 Paris Cedex 13, France}
\altaffiltext{4}{CEA, Centre de Saclay, Irfu/SPP, F-91191 Gif-sur-Yvette, France}
\altaffiltext{5}{Department of Astronomy, University of Washington, 3910 15th Ave. NE, Seattle, WA 98195, USA}
\altaffiltext{6}{Cutler Group, LP, 101 Montgomery St., Suite 700, San Francisco, CA 94104, USA}
\altaffiltext{7}{Department of Astronomy, University of Wisconsin, 475 North Charter St., Madison, WI 53706, USA}
\altaffiltext{8}{Harvard-Smithsonian Center for Astrophysics, 60 Garden St., Cambridge, MA 02138, USA}
\altaffiltext{9}{Department of Astrophysical Sciences, Peyton Hall, Princeton University, 4 Ivy Ln., Princeton, NJ 08544, USA}
\altaffiltext{10}{Institute of Cosmology and Gravitation, University of Portsmouth, Portsmouth PO1 3FX}
\altaffiltext{11}{Center for Cosmology and Particle Physics, New York University, 4 Washington Pl., New York, NY 10003, USA}
\altaffiltext{12}{Department of Physics and Astronomy, University of Wyoming, Laramie, WY 82071, USA}
\altaffiltext{13}{Department of Physics, Yale University, 260 Whitney Ave, New Haven, CT 06520, USA}
\altaffiltext{14}{Institut d'Astrophysique de Paris, UPMC-CNRS, UMR7095, 98bis blvd.\ Arago, 75014 Paris, France}
\altaffiltext{15}{Departamento de Astronom\'{i}a, Universidad de Chile, Casilla 36-D, Santiago, Chile}
\altaffiltext{16}{UCO/Lick Observatory, University of California, Santa Cruz, 1156 High St., Santa Cruz, CA 95064, USA}
\altaffiltext{17}{Department of Astronomy and Astrophysics, The Pennsylvania State University, University Park, PA 16802, USA}
\altaffiltext{18}{Institute for Gravitation and the Cosmos, The Pennsylvania State University, University Park, PA 16802, USA}
\altaffiltext{19}{Department of Astronomy, Yale University, 260 Whitney Ave., New Haven, CT 06520, USA}
\altaffiltext{20}{Department of Physics and Astronomy, University of Pittsburgh, 100 Allen Hall, 3941 O'Hara St., Pittsburgh, PA 15260, USA}

\begin{abstract}
We describe the automated spectral classification, redshift
determination, and parameter measurement
pipeline in use for the Baryon Oscillation Spectroscopic Survey (BOSS)
of the Sloan Digital Sky Survey III (SDSS-III)
as of the survey's Ninth Data Release (DR9), encompassing
831,000 moderate-resolution optical spectra.
We give a review of the algorithms employed, and describe
the changes to the pipeline
that have been implemented for BOSS relative to previous SDSS-I/II versions,
including new sets of stellar, galaxy, and quasar redshift templates.
For the color-selected ``CMASS'' sample of massive galaxies
at redshift $0.4 \la z \la 0.8$ targeted by BOSS for
the purposes of large-scale cosmological measurements,
the pipeline achieves an automated classification success rate
of 98.7\% and confirms 95.4\% of unique CMASS targets as galaxies
(with the balance being mostly M stars).  Based on visual inspections of a subset
of BOSS galaxies, we find that approximately 0.2\% of
confidently reported CMASS sample classifications and
redshifts are incorrect,
and about 0.4\% of all CMASS spectra are objects
unclassified by the current algorithm which are potentially recoverable.
The BOSS pipeline confirms that $\sim$51.5\%
of the quasar targets have quasar spectra,
with the balance mainly consisting of stars
and low signal-to-noise spectra.
Statistical (as opposed to systematic)
redshift errors propagated from photon noise
are typically a few tens of km\,s$^{-1}$ for both galaxies
and quasars, with a significant tail to a few hundreds of
km\,s$^{-1}$ for quasars.
We test the accuracy of these statistical redshift error estimates using repeat
observations, finding them underestimated
by a factor of 1.19 to 1.34 for galaxies, and
by a factor of 2 for quasars.
We assess the impact of sky-subtraction quality,
signal-to-noise ratio, and other factors on galaxy redshift success.
Finally, we document known issues with the BOSS DR9 spectroscopic data set,
and describe directions of ongoing development.
\end{abstract}


\slugcomment{Submitted to The Astronomical Journal}

\maketitle

\section{Introduction}

The Sloan Digital Sky Survey III (SDSS-III, \citealt{Eisenstein11})
is the third phase of the SDSS \citep{York00}.\footnote{Throughout
this paper, we will refer to the earlier SDSS phases collectively as SDSS-I/II\@.}
Within the SDSS-III, the Baryon Oscillation Spectroscopic Survey (BOSS, \citealt{Dawson12})
is currently mapping a larger volume of the universe
than any previous spectroscopic survey.
The Ninth Data Release of the SDSS-III (DR9, \citealt{Nine12},
released publicly on 2012 July 31)
is the first SDSS-III data release to include BOSS spectroscopic data, and
comprises good observations of 831 unique plate-pluggings
of 813 unique tilings (plates worth of targets) on the sky.
Each plate delivers simultaneous spectroscopic
observations of 1000 lines of sight with optical fibers that feed
a pair of two-arm spectrographs, giving a total
of 831,000 BOSS DR9 spectra.

The main science goal of BOSS is to trace the large-scale mass
structure of the universe using massive galaxies and quasar Ly$\alpha$
absorption systems, in order to measure the length scale of the
``baryon acoustic oscillation'' feature in the spatial correlation
function of these objects \citep[e.g.,][]{Eisenstein05}, and thereby to constrain the
nature of the dark energy that drives the accelerated expansion of
the present-day universe.  To meet this goal, the BOSS project
has specified a series of scientific requirements, including:
(1) an RMS galaxy redshift precision better than
300 km\,s$^{-1}$; (2) a galaxy redshift success rate of at least 94\%, including
both targeting inefficiency and spectroscopic redshift failure;
(3) a catastrophic galaxy redshift error rate of less than 1\%; and
(4) spectroscopic confirmation of at least 15 quasars at $2.2 < z < 3.5$
per degree$^2$ from among no more than 40 targets per degree$^2$.
To satisfy these requirements within such a large
survey, automated spectroscopic calibration, extraction, classification,
and redshift measurement methods are essential.

This paper, one of a series of technical papers
describing SDSS-III DR9 in general and the BOSS
data set in particular, presents the automated classification
and redshift measurement software for the main galaxy
and quasar target samples implemented for the BOSS project.
This software is written in the IDL
language, and is titled \texttt{idlspec2d}.
Earlier versions of this code were used to
analyze SDSS-I/II data (see \citealt{Aihara11}),
alongside the complementary and independently developed
pipeline software \texttt{spectro1d} (see \citealt{Subbarao02}
and \citealt{Adelman06});
for the BOSS project, the \texttt{idlspec2d} software
has been adopted as the primary code, due to its
robust error estimation methods and its tight integration of
redshift measurement and classification with the lower-level
operations of raw data calibration and extraction.
The code has also been upgraded with new redshift-measurement
templates and several new algorithms in order
to meet the scientific requirements of the BOSS project.
The tagged software version \texttt{v5\_4\_45} was used to
process all BOSS spectroscopic
data for DR9\footnote{The DR9 tagged version of \texttt{idlspec2d}
can be obtained at
\texttt{www.sdss3.org/svn/repo/idlspec2d/tags/v5\_4\_45/}.},
and the classification and
redshift results delivered by this code have
been used for recently published BOSS DR9-sample cosmological analyses
\citep{Anderson12,Manera12,Nuza12,Reid12,RossA12,Sanchez12,Tojeiro12}.
An overview of the BOSS project, including experimental
design, scientific goals, observational
operations, and ancillary programs, is given in \citet{Dawson12}.
A description of the \texttt{idlspec2d} calibration
and extraction methods which transform raw CCD pixel data
into one-dimensional object spectra
will be presented in \citet{Schlegel12}.

The organization of this paper is as follows.
Section~\ref{sec:data} presents an overview of
the spectroscopic data sample of BOSS DR9.
Section~\ref{sec:pipeline} describes
the classification and redshift pipeline
algorithms and procedures, including the core
redshifting algorithm (\S\ref{subsec:zmeasure}),
special classification handling for the
galaxy target samples (\S\ref{subsec:z_noqso}),
measured spectroscopic
parameters (\S\ref{subsec:params}),
and output files (\S\ref{subsec:outfiles}).
Section~\ref{sec:templates} describes the
templates constructed for the automated
spectroscopic identification and redshift
analysis of BOSS galaxies (\S\ref{subsec:galtemp}),
quasars (\S\ref{subsec:qsotemp}), and stars (\S\ref{subsec:startemp}).
Section~\ref{sec:performance} analyzes the completeness,
purity, accuracy, and precision of the
samples classified and measured by the \texttt{idlspec2d} pipeline.
Section~\ref{sec:issues} documents known issues in the
DR9 release of BOSS data, and \S\ref{sec:summary}
provides a summary and conclusions.

\section{Data Overview}
\label{sec:data}

The main BOSS survey program consists of two galaxy target
samples \citep{Padmanabhan12} and a quasar target sample
including both color-selected candidates and known quasars
\citep{Bovy11,Kirkpatrick11,RossN12}.
The galaxy samples are designated CMASS (for ``constant mass'')
and LOWZ (for ``low-redshift'').
The LOWZ galaxy sample is composed of massive red galaxies spanning the
redshift range $0.15 \la z \la 0.4$.  The CMASS galaxy sample is composed of massive
galaxies spanning the redshift range $0.4 \la z \la 0.7$.  Both samples are
color-selected to provide near-uniform sampling over the combined volume.
The faintest galaxies are at $r=19.5$ for LOWZ and $i=19.9$ for CMASS\@.
Colors and magnitudes for the galaxy selection cuts
are corrected for Galactic extinction
using \citet{Schlegel98} dust maps.
The BOSS quasar sample is selected to recover as many objects as possible in
the redshift range $2.2 < z < 3.5$ for the purposes of measuring the
3D structure in the Ly$\alpha$ forest.
A variety of selection algorithms are employed
to select the quasar sample,
which lies close to the color locus of F stars.
The faint-end magnitude limits of the quasar target sample are
extinction-corrected PSF magnitudes of
$g=22$ and $r=21.85$.

\begin{figure*}[t]
\epsscale{1.1}
\plotone{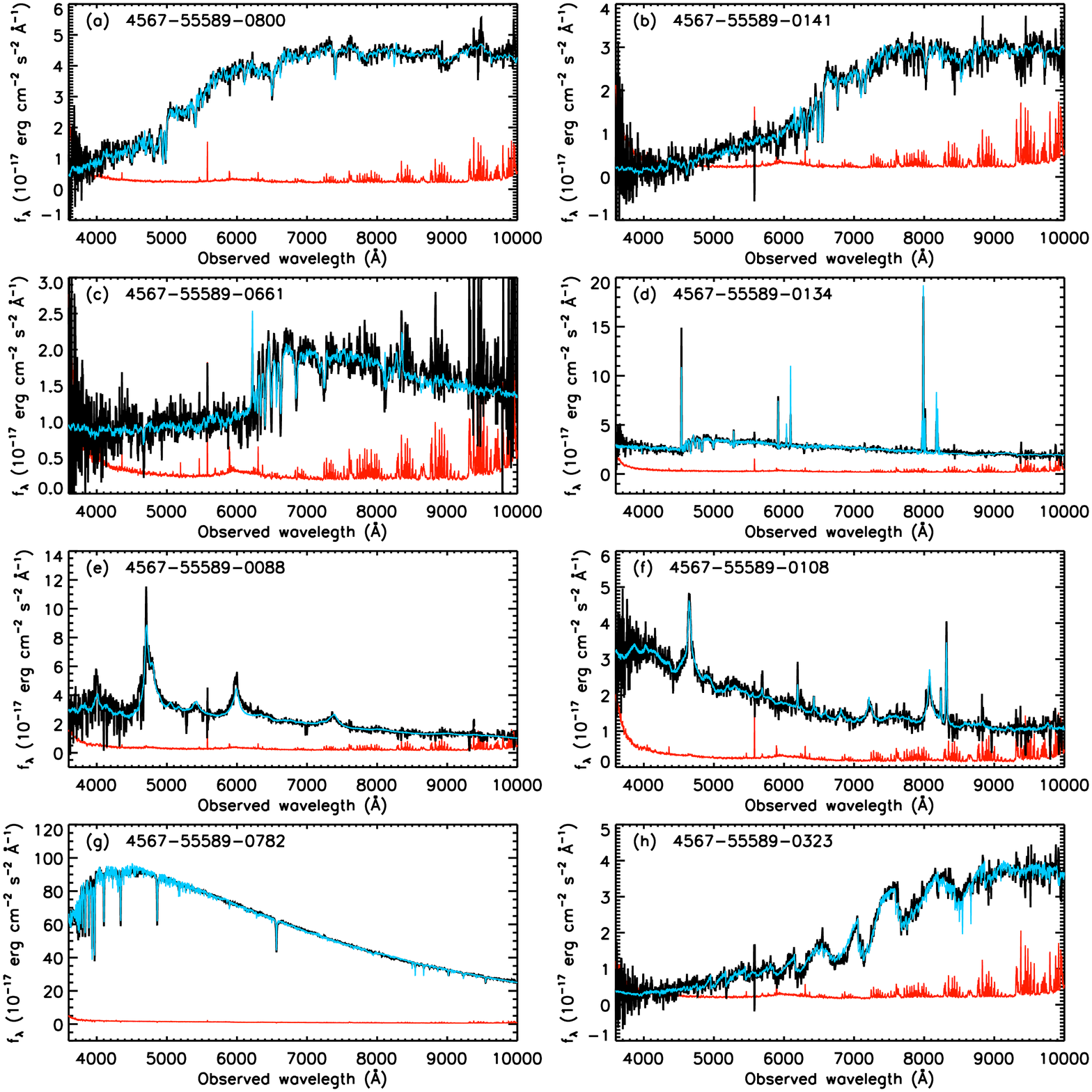}
\caption{\label{fig:mosaic}
Mosaic of representative BOSS spectra, with a resolution
of $R\approx 2000$.  Black lines show data (smoothed over a 5-pixel window),
cyan lines show best-fit redshift/classification model, and red lines show 1-$\sigma$ noise level
estimated by the extraction pipeline.  Spectra are labeled
by \texttt{PLATE-MJD-FIBERID}.  Individual objects are: (a) redshift $z = 0.256$ LOWZ galaxy;
(b) redshift $z = 0.649$ CMASS galaxy; (c) redshift $z = 0.669$ CMASS galaxy with post-starburst continuum;
(d) redshift $z = 0.217$ starburst galaxy (from QSO target sample);
(e) redshift $z = 2.873$ quasar; (f) redshift $z = 0.661$ quasar;
(g) spectrophotometric standard star;
(h) M star (from CMASS target sample).}
\end{figure*}

A summary of the DR9 BOSS spectroscopic data set
(observed between 2009 December and 2011 July) is
given in Table~\ref{table:dr9summary}, along with performance
metrics that will be discussed in detail further below.
Representative BOSS survey spectra are shown in Figure~\ref{fig:mosaic}.
The automated classification and measurement software described
here is applied to all spectra obtained by the BOSS
spectrographs \citep{Smee12}, including spectra targeted under
ancillary programs described in \citet{Dawson12}.
In this work we focus on the analysis of the main
BOSS galaxy and quasar survey targets, since
the performance on these samples is the primary scientific
driver of the design, development, and verification
of the pipeline.

\begin{table}[t]
\begin{center}
\caption{\label{table:dr9summary} BOSS DR9 summary spectrum totals}
\begin{tabular}{lr}
\hline \hline
Item & Number \\
\hline
Plate pluggings &          831 \\
Unique plates &          819 \\
Unique tiles (plates worth of targets) &          813 \\
Spectra &       831000 \\
Effective spectra\tablenotemark{1} &       829073 \\
Unique spectra &       763425 \\
CMASS sample spectra &       353691 \\
Unique CMASS spectra &       324198 \\
Unique CMASS with \texttt{ZWARNING\_NOQSO == 0}\tablenotemark{2} &       320031 \\
Unique CMASS that are galaxies &       309307 \\
LOWZ sample spectra &       111347 \\
Unique LOWZ spectra &       103729 \\
Unique LOWZ with \texttt{ZWARNING\_NOQSO == 0}\tablenotemark{2} &       103610 \\
Unique LOWZ that are galaxies &       102890 \\
CMASS \&\& LOWZ sample spectra &         3201 \\
Unique CMASS \&\& LOWZ spectra &         2990 \\
Unique CMASS \&\& LOWZ with \texttt{ZWARNING\_NOQSO == 0}\tablenotemark{2} &         2976 \\
Unique CMASS \&\& LOWZ that are galaxies &         2935 \\
Quasar sample spectra\tablenotemark{3} &       166034 \\
Unique quasar sample spectra &       154433 \\
Unique quasar sample with \texttt{ZWARNING == 0}\tablenotemark{4} &       122488 \\
Unique quasar sample spectra that are quasars &        79570 \\
Number of above with $2.2 \le z \le 3.5$ &        51903 \\
Unique quasar sample scanned visually &       154173 \\
Visual $2.2 \le z \le 3.5$ quasars missed by pipeline &          895 \\
Pipeline $2.2 \le z \le 3.5$ QSOs with visual disagreement\tablenotemark{5} &          327 \\
Sky spectra &        78573 \\
Unique sky-spectrum lines of sight &        75850 \\
Spectrophotometric standard star spectra &        16905 \\
Unique standard star spectra &        14915 \\
Ancillary program spectra &        32381 \\
Unique ancillary target spectra &        28968 \\
Other spectra (commissioning, calibration, etc.) &        74620 \\
Unique other spectra &        65461 \\
\hline
\end{tabular}
\tablenotetext{1}{Excludes unplugged fibers and spectra falling on bad CCD columns.}
\tablenotetext{2}{\texttt{ZWARNING\_NOQSO == 0} indicates a confident spectroscopic classification and redshift measurement for galaxy targets (see \S\ref{subsec:zmeasure} and \S\ref{subsec:z_noqso}).}
\tablenotetext{3}{``Quasar targets'' tabulated here are from the main survey quasar sample, and exclude any ancillary and calibration quasar targets.}
\tablenotetext{4}{\texttt{ZWARNING == 0} indicates a confident spectroscopic classification and redshift measurement for quasar targets (see \S\ref{subsec:zmeasure}).}
\tablenotetext{5}{``Visual disagreement'' is either $|\Delta z| > 0.05$ between pipeline and visual inspections, or absence of confident visual classification \& redshift.}
\end{center}
\end{table}

For the purposes of this paper, we define
the samples of unique LOWZ and CMASS spectra
according to the following cuts:
\begin{enumerate}
\item Selected by the appropriate sample color cuts
(encoded by bit 0 of the \texttt{BOSS\_TARGET1} mask for the LOWZ
sample, and by bit 1 of that mask for the CMASS sample.)
The LOWZ and CMASS samples are not mutually exclusive,
although they are mostly non-overlapping.
\item \label{item:withdata} Observed with a spectroscopic fiber that is
well plugged, successfully mapped to the target object,
and not affected by bad CCD columns
that remove a large fraction of the wavelength coverage.
These conditions are reported via bits 1 and 7 of the
\texttt{ZWARNING} bitmask described in \S\ref{subsec:zmeasure}.
\item Apparent (not extinction-corrected)
$i$-band magnitude less than 21.5
within a 2$\arcsec$-diameter circular
aperture, corresponding to
the angular size of the BOSS fibers.
This criterion excludes low surface-brightness targets for which the
spectroscopic signal-to-noise ratio (S/N) becomes unacceptably low
for nominal survey exposure times of 60 minutes in good conditions.
\item \label{item:specprimary} Best single observation within the
survey data set, for the case of multiply observed
spectra.  This designation is described in
\S\ref{subsec:outfiles}.
\end{enumerate}

The sample of unique BOSS quasar spectra for the current work is defined
according to the following cuts:
\begin{enumerate}
\item Selected from one of the four categories of
known quasars with redshifts optimal for Ly$\alpha$
forest analysis (bit 12 of the \texttt{BOSS\_TARGET1} mask),
quasars selected from the FIRST
survey \citep[bit 18,][]{Becker95}, and candidates
from the BOSS ``Core'' and ``Bonus'' quasar candidate selection algorithms
(bits 40 and 41 respectively: see \citealt{RossN12}).
\item Plugged, mapped, and well-covered in wavelength
(as for Item~\ref{item:withdata} of the previous list
for galaxy targets).
\item Best single observation within the survey data
set (as for Item~\ref{item:specprimary} of the previous list).
\end{enumerate}

\section{Pipeline Overview}
\label{sec:pipeline}

Imaging and spectroscopic data for the BOSS Survey are obtained
with the 2.5-m Sloan Telescope at Apache Point Observatory
\citep{Gunn06}, first with the imaging camera \citep{Gunn98}
and then with an upgraded (relative to SDSS-I/II)
spectrograph system capable of obtaining
1000 spectra simultaneously using optical fibers plugged
into a drilled aluminum focal plate and feeding two double-arm
spectrographs.  The characteristics of this instrument
are summarized in Table~\ref{table:spectrograph}, and described
in detail by \citet{Smee12}.  The outputs of the fibers feeding each spectrograph
are arrayed linearly along a ``slit-head'' and numbered within
the spectroscopic pipeline by the sequential index \texttt{FIBERID},
which by convention runs from 1--500 in spectrograph 1 and from 501-1000
in spectrograph 2.  A unique physical target plate is specified by the
\texttt{PLATE} identifier.  Since the same plate can be plugged
and observed on multiple occasions, with different mappings between fibers
and target holes, the modified Julian date of a unique plugging is
tracked as well via the \texttt{MJD} parameter.  Together, the
combination of \texttt{PLATE}, \texttt{MJD}, and \texttt{FIBERID}
constitute a unique identifier for a BOSS spectrum (as was also the
case for SDSS-I/II spectra).  Each plugging is observed with multiple
exposures which are exactly 15 minutes each in length and
can be distributed across more than one night of observation.
Typically four to six exposures are required to attain sufficient
S/N per pixel at a fiducial magnitude.
All good data from a unique plugging are co-added together during the spectroscopic
data reduction process.  Data from different pluggings are never
combined together.

\begin{table}
\begin{center}
\caption{\label{table:spectrograph} BOSS spectrograph system characteristics}
\begin{tabular}{ll}
\hline \hline
Parameter & Value \\
\hline
On-sky field of view & 3$^{\circ}$ diameter \\
Fiber aperture & 2$\arcsec$ diameter \\
Multiplex capability & 1,000 objects \\
Wavelength coverage & 3,600\,\AA\,$\la \lambda \la$\,10,400\,\AA\ \\
Spectral resolution & $\lambda / \Delta \lambda \approx$\,2,000 \\
\hline
\end{tabular}
\end{center}
\end{table}

The wavelength calibration, extraction, sky subtraction,
flux calibration, and co-addition of
BOSS spectra from raw CCD pixel data are
described in \citet{Schlegel12}.  The extraction
implementation is a variation of the
optimal-extraction algorithm described by \citet{Hewett85} and
\citet{Horne86}, including a forward-modeling solution that de-blends the
cross-talk between neighboring fibers on the CCD\@.
(A similar approach is described in \citealt{Sandin10}.)
The outputs of this ``two-dimensional'' pipeline software are
stored on a plate-by-plate basis for sets of 1000 spectra in the
multi-extension ``\texttt{spPlate}'' FITS files, which are
the inputs to the ``one-dimensional'' (1D) pipeline software described
in this work.  The full contents of the \texttt{spPlate} files are described
in detail in the online data model\footnote{\texttt{http://www.sdss3.org/dr9/}};
for the purposes of the redshift measurement and classification pipeline,
the most important products are:
\begin{enumerate}
\item Wavelength-calibrated, sky-subtracted, flux-calibrated, and
co-added object spectra, rebinned onto a uniform baseline
of $\Delta \log_{10} \lambda = 10^{-4}$ (about 69\,km\,s$^{-1}$) per pixel.
\item Statistical error-estimate vectors for each spectrum
(expressed as inverse variance) incorporating contributions
from photon noise, CCD read noise, and sky-subtraction error.
\item Mask vectors for each spectrum identifying pixels where
warning conditions exist in either any (\texttt{ORMASK})
or all (\texttt{ANDMASK}) of the spectra
contributing to the final co-added spectrum.
\end{enumerate}

\subsection{Redshift measurement and classification}
\label{subsec:zmeasure}

The BOSS spectral classification and redshift-finding analysis
is approached
as a $\chi^2$ minimization problem.  Linear fits are made to each observed
spectrum using multiple templates and combinations of templates
evaluated for all allowed redshifts, and the global minimum-$\chi^2$
solution is adopted for the output classification
and redshift.  This approach requires the
spectra and their errors to be well-understood, and requires
the template spectra to sufficiently span the observed space.
Both these conditions are satisfied for the BOSS galaxy and quasar targets,
resulting in accurate redshift fits and enabling a quantitative assessment
of the confidence of those fits.  By performing a
statistically objective analysis, confident redshifts are obtained
even for data at lower S/N where manual inspection may fail.

The basic outputs of the redshift determination and classification
algorithm described in this section
are the measured redshift \texttt{Z}, its
associated 1-sigma statistical error \texttt{Z\_ERR},
a classification category \texttt{CLASS} (either
\texttt{"GALAXY"}, \texttt{"QSO"} for quasar, or \texttt{"STAR"}),
and a confidence flag \texttt{ZWARNING} that is
zero for confident measurements and non-zero otherwise.
Section~\ref{subsec:z_noqso} describes special variations
on these outputs that are recommended for use with the BOSS
LOWZ and CMASS galaxy sample spectra.

\begin{figure}
\epsscale{1.15}
\plotone{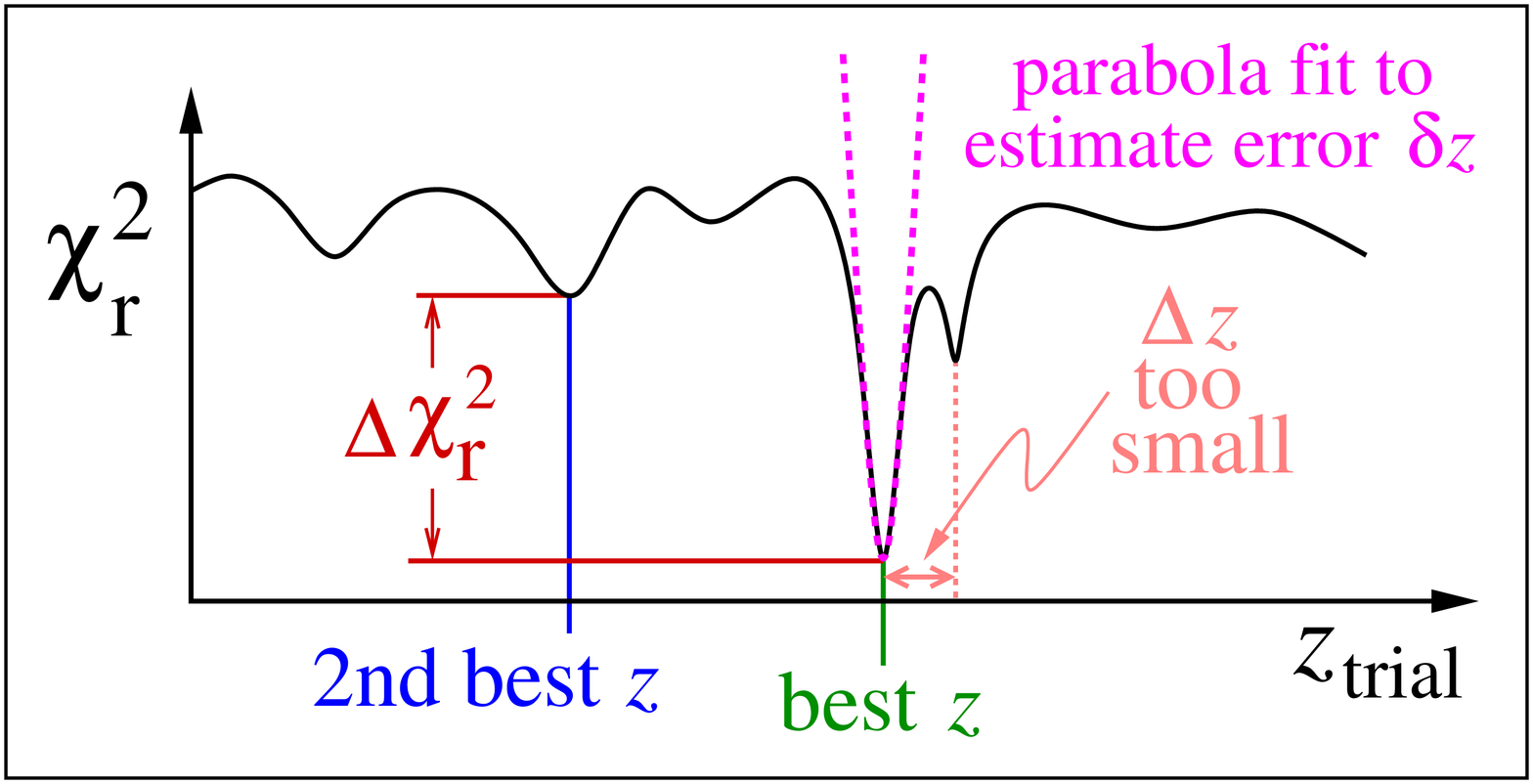}
\caption{\label{fig:chi2fit}
Schematic illustration of the \texttt{idlspec2d} redshift
measurement algorithm.
The reduced $\chi^2$ curve as a function of
redshift \textit{(black curve)} is determined
from the best-fit linear combination of template basis
spectra at each trial redshift value.
The best redshift is defined by the location of the
global minimum \textit{(green)}.  Subsidiary minima separated
by less than 1000\,km\,s$^{-1}$ are not considered
to be separate \textit{(pink)}.  The curvature of a parabolic fit
to the $\chi_r^2$ curve at the global minimum \textit{(magenta)}
is used to determine the best-fit redshift
error estimate.  The second-best redshift fit is determined
by the location of the second-lowest well-separated $\chi_r^2$
minimum \textit{(blue)}.  The difference $\Delta \chi_r^2$
\textit{(red)} between best and
second-best redshifts is used to assign
confidence in the measurement, as described in the text.}
\end{figure}

The least-squares minimization is performed by comparison of each
spectrum to a full range of templates spanning galaxies, quasars, and stars.
A range of redshifts is explored, with trial redshifts spaced
every pixel (69\,km\,s$^{-1}$) for most templates and spaced
by four pixels (276\,km\,s$^{-1}$) for quasar templates.
At each redshift the spectrum is fit with an error-weighted least-squares
linear combination of redshifted template ``eigenspectra''
in combination with a low-order polynomial.
The polynomial terms absorb Galactic extinction, intrinsic
extinction, and residual spectro-photometric calibration
errors (typically at the 10\% level)
that are not fully spanned by the eigenspectra.
The template basis sets are derived from rest-frame
principal-component analyses (PCA) of
training samples of galaxies, quasars, and cataclysmic variable stars,
and from a set of archetype spectra for other types
of stars.  CV stars are handled as a separate class
from other stars due
to their significant range of emission-line strengths.
The construction of these basis sets is described in detail
in \S\ref{sec:templates} below.
This best-fit combination model gives a $\chi^2$ value for that
trial redshift, and these values define a $\chi^2 (z)$ curve
when computed across the redshift range under consideration.
To facilitate comparison between template
classes with differing numbers of basis vectors, these
$\chi^2$ values are analyzed in reduced form $\chi^2_r$,
i.e., divided by the number of degrees of freedom.
In practice this is nearly equivalent to working in
terms of $\chi^2$ for any given spectrum, as the number of
pixels ($\sim$\,4500) greatly exceeds the number of free parameters in
all fits.  The best redshifts for a particular
class under consideration
are defined by the locations of the lowest minima in
the $\chi^2_r$ curve, where that curve is fit by a quadratic
function using the five points nearest the minimum (11 points for quasars).
The initial quasar fits where the trial redshifts are spaced every four pixels
are re-fit near the five best fits with redshifts spaced every pixel.
This two-step fitting for the quasars is done for computational efficiency,
since most of the computational time is spent evaluating quasar fits
which are performed on all targets.
The statistical error on the final redshift is evaluated at
the location of the minimum of the $\chi^2$ curve as
the change in redshift $\pm \delta z$
for which $\chi^2$ increases by one above the minimum value.
Noise in the spectra can result in multiple local minima in
the neighborhood of the global minimum that are not significant.
These are typically separated by a few pixels,
or $\sim$200\,km\,s$^{-1}$.  For all template fits, we collapse
minima separated by less than 1000\,km\,s$^{-1}$ to a single minimum.
Solutions with separations exceeding  1000\,km\,s$^{-1}$
must be explicitly evaluated since they represent
catastrophic redshift failures for BOSS galaxies
if they are statistically indistinguishable from one another
(see panel ``h'' of Figure~\ref{fig:crappo} further below).
The redshift-finding procedure is shown schematically in Figure~\ref{fig:chi2fit}.
(See also \citealt{Glazebrook98}.)

This core algorithm is applied within the pipeline according
to the following sequence:
\begin{enumerate}
\item Read the spectrum, error estimates,
and mask vectors for a single spectroscopic plate
from the \texttt{spPlate} file.
\item Mask pixels outside the range 3600\,\AA--10400\,\AA,
pixels at wavelengths
where the typical sky-subtraction model residuals are more
than 3-sigma worse than the errors
expected from a Poisson model in
any sub-exposure (\texttt{BADSKYCHI} set in the \texttt{ORMASK}
vector output by the reduction software; \citealt{Schlegel12}),
pixels where the sky brightness is in excess of the
extracted object flux plus ten times its statistical
error in all sub-exposures
(\texttt{BRIGHTSKY} set in the \texttt{ANDMASK}), and
pixels with negative flux at greater than 10-$\sigma$
significance.
\item Find the best five galaxy redshifts between
$z = -0.01$ and $z = 1.0$, using a rest-frame template
basis of four eigenspectra (\S\ref{subsec:galtemp}).
\item Find the best five quasar redshifts between
$z=0.0033$ and $z = 7.0$, using a rest-frame template
basis of four eigenspectra (\S\ref{subsec:qsotemp}).
\item Find the best single redshift for each of 123 sub-classes
of star from $-$1200\,km\,s$^{-1}$ to $+$1200\,km\,s$^{-1}$,
using a single rest-frame archetype spectrum
for each one (\S\ref{subsec:startemp}).
\item Find the best single cataclysmic variable star redshift
from $-$1000\,km\,s$^{-1}$ to $+$1000\,km\,s$^{-1}$,
using a rest-frame template basis of three eigenspectra
(in order to capture emission-line strength variations,
\S\ref{subsec:startemp}.)
\item Sort all redshifts and classifications together
by ascending $\chi_r^2$, and assign the best
spectroscopic redshift
and classification from among \texttt{GALAXY},
\texttt{QSO} (quasar), and \texttt{STAR} (including CV)
based on the overall minimum
$\chi_r^2$ across all classes and redshifts.
\item Compare the change in $\chi_r^2$ between
the best classification and redshift and the
next-best classification and redshift
with a velocity difference greater than 1000\,km/s$^{-1}$,
and assign a low-confidence ``\texttt{ZWARNING}''
flag if this difference is either less than 0.01 in absolute terms
or less than 0.01 times the overall minimum $\chi_r^2$ value.
For the $\sim$\,4500
degrees of freedom typical of a BOSS spectrum,
the absolute threshold of
$\Delta \chi_r^2 = 0.01$ corresponds
to $\Delta \chi^2 \approx 45$ (naively interpreted
as 6.7-sigma).
The relative requirement on $\Delta \chi_r^2$
serves to make the statistical
confidence threshold progressively more
conservative at higher S/N levels where the
redshift templates fits are worse in an absolute sense
but nevertheless have greater distinguishing
power among multiple hypotheses.
\end{enumerate}

The threshold value of $\Delta \chi_r^2 > 0.01$
used to assign confidence to the classifications
is empirically determined.
The threshold could be lowered further to recover more redshifts but at the
cost of more mis-classifications and incorrect redshifts.
Tests on repeat CMASS sample data show that
decreasing the threshold to 0.008 (0.005) would
increase redshift completeness by 0.3\% (0.6\%),
with 8\% (16\%) of the added measurements
being incorrect.
(A full analysis of BOSS galaxy redshift
completeness and purity is given in \S\ref{subsec:galcomp}.)
An additional test is
made possible by the nearly 80,000 blank BOSS sky spectra in DR9.
Among these, 2\% fit a template with a confidence
$\Delta \chi_r^2 > 0.01$, implying they would be assigned a confident redshift
in the absence of prior knowledge of their status as blank sky spectra.
(Although a small fraction of these are in fact real
objects detected spectroscopically in the sky fibers.)

As discussed above, the condition \texttt{ZWARNING}\,$= 0$
designates that the BOSS pipeline has determined a confident
classification and redshift for a spectrum.
The primary source of \texttt{ZWARNING}\,$\ne 0$ spectra
is the $\Delta \chi_r^2$ threshold.  However,
several other flags are also encoded bit-wise in
the \texttt{ZWARNING} mask, as
documented in Table~\ref{table:zwarning}.
The definitions of the \texttt{ZWARNING}
mask-bits in BOSS are identical to their definitions
in SDSS-I/II\@.  Two of the bits have been disabled
for BOSS DR9, and are only retained for historical consistency:
(1) the \texttt{NEGATIVE\_MODEL} bit, which
was previously triggered by stellar model fits with negative amplitudes,
which are now disallowed entirely; and (2)
the \texttt{MANY\_OUTLIERS} bit, which was found to flag too many
good, high-S/N quasar redshifts in BOSS\@.

\begin{table*}[t]
\begin{center}
\caption{\label{table:zwarning} BOSS DR9 redshift and classification warning flags (ZWARNING)}
\begin{tabular}{lll}
\hline \hline
Bit & Name & Definition \\
\hline
0 & \texttt{SKY} & Sky fiber \\
1 & \texttt{LITTLE\_COVERAGE} & Insufficient wavelength coverage \\
2 & \texttt{SMALL\_DELTA\_CHI2} & $\Delta \chi_r^2$ between best and second-best fit is less than 0.01 (or 0.01 $\times$ the minimum $\chi_r^2$) \\
3 & \texttt{NEGATIVE\_MODEL} & Synthetic spectrum negative, \textbf{disabled for BOSS DR9} \\
4 & \texttt{MANY\_OUTLIERS} & More than 5\% of points above 5-$\sigma$ from synthetic spectrum, \textbf{disabled for BOSS DR9} \\
5 & \texttt{Z\_FITLIMIT} & $\chi^2$ minimum for best model is at the edge of the redshift range \\
6 & \texttt{NEGATIVE\_EMISSION} & Negative emission in a quasar line at 3-$\sigma$ significance or greater (see \S\ref{subsec:params}) \\
7 & \texttt{UNPLUGGED} & Broken or unplugged fiber \\
\hline
\end{tabular}
\end{center}
\end{table*}

Table~\ref{table:templates} summarizes the number of
PCA template and polynomial degrees of freedom associated with each spectroscopic
object class, along with the name of the file containing the
template basis within the \texttt{idlspec2d/v5\_4\_45}
software product.
In most cases, the number of PCA templates and
number of polynomial terms used
in the redshift and classification analysis match
the SDSS-I/II \texttt{idlspec2d} values.  The one exception is the number
of CV star templates, which has decreased from four in
SDSS-I/II to three in BOSS due to
a smaller available training sample at the time the DR9
code version was frozen.  For all target classes
we have verified that the choices are nearly optimal by performing
tests of the completeness and purity of classification and
redshift measurement (relative to visually inspected subsets) as
a function of the size of the PCA and polynomial
basis.
As can be expected, increasing the number of PCA and polynomial terms
used for the modeling of a particular class increases both the
completeness and the \textit{im}purity of the resulting sample for that class.
Increases in impurity arise from both catastrophic mis-classification
and catastrophic redshift error, with the former decreasing the completeness
of other classes.
Each spectrum in the survey is fitted with all classes
of objects in order to determine a spectroscopic
redshift and classification that is
independent of photometric data and targeting information
(but see \S\ref{subsec:z_noqso} below).

\begin{table}[b]
\begin{center}
\caption{\label{table:templates} Summary of BOSS redshift \& classification degrees of freedom}
\begin{tabular}{lcc}
\hline \hline
 ~  & Template & $N_{\mathrm{temp}}$, $N_{\mathrm{poly}}$ \\
Class & Filename & Per Fit\tablenotemark{1} \\
\hline
\texttt{GALAXY} & \texttt{spEigenGal-55740.fits} & 4, 3 \\
\texttt{QSO} & \texttt{spEigenQSO-55732.fits} & 4, 3 \\
\texttt{STAR}\tablenotemark{2} & \texttt{spEigenStar-55734.fits} & 1, 4 \\
\texttt{STAR} (CV) & \texttt{spEigenCVstar-55734.fits} & 3, 3 \\
\hline
\end{tabular}
\tablenotetext{1}{$N_{\mathrm{temp}}$ is number of basis templates
per fit; $N_{\mathrm{poly}}$ is number of additive polynomial
background terms used in addition to template basis
in the fit.}
\tablenotetext{2}{There are a total of 123 non-CV stellar subtypes
considered, but each trial fit only includes a single
template.}
\end{center}
\end{table}

\subsection{Special galaxy target handling}
\label{subsec:z_noqso}

The implementation of the \texttt{idlspec2d} redshift code is designed
to meet the BOSS scientific requirements on redshift success rates,
as discussed in the Introduction.
The original SDSS-I/II code
operated on spectra alone, without imposing classification
or redshift priors based on photometric data or other
targeting information.  At the S/N typical of SDSS-I/II
spectra, this technique proved highly successful, resulting in
a redshift success rate better than 99\% for the main galaxy sample and
a negligible incidence of catastrophic errors.
For the BOSS galaxy samples, however, some prior information from the
targeting photometric catalog is needed to achieve the required
redshift success rate.  Specifically, we have found in practice
that the LOWZ and CMASS targets can be
galaxies, stars, or superpositions of the two, but are
almost never quasars (however, see Item~\ref{item:type2}
in \S\ref{sec:issues}.)
Without using any prior information, the redshift code
produces an excess of erroneous quasar classifications for CMASS targets due to
unphysical quasar basis-plus-polynomial combinations yielding
the global minimum-$\chi^2$ (see panel ``a'' in Figure~\ref{fig:crappo}
further below.)

To remedy this, the adopted BOSS
survey values for the redshifts of LOWZ and CMASS
galaxy targets are taken from
the parameters \texttt{Z\_NOQSO} and \texttt{CLASS\_NOQSO},
together with the associated statistical error estimates
\texttt{Z\_ERR\_NOQSO} and confidence flags \texttt{ZWARNING\_NOQSO},
which represent the
best-fit redshift and classification
determined through the procedure
described in \S\ref{subsec:zmeasure}, but \textit{excluding the consideration of quasar
template fits.}  This effectively imposes the red-color and
extended-image priors of the galaxy target sample
over the blue-color and point-like image priors of
the quasar target sample.
The \texttt{SMALL\_DELTA\_CHI2} bit for the \texttt{ZWARNING\_NOQSO}
mask is set only on the absolute criterion
of $\Delta \chi_r^2 < 0.01$ relative
to the next-best non-quasar model
(i.e., with no relative $\Delta \chi_r^2$ cut).
We recommend the use of these ``\texttt{NOQSO}'' quantities for statistical
analyses of the BOSS galaxy samples.
The parameter \texttt{Z}
and its associated values are also retained and reported
for consistency
with the original SDSS-I/II approach, representing the global
minimum-$\chi^2$ redshift inclusive of all spectral template classes.

\subsection{Parameter measurements}
\label{subsec:params}

The primary outputs of the analysis code described
in this work are the classification, redshift, redshift error,
and best template-based model fit to each spectrum.
The code also measures a number of parameters
assuming the best-fit classification and redshift.
Specifically: stellar velocity dispersions are measured for
objects classified as galaxies; emission-line parameters are measured for
all objects; and supplemental stellar sub-classifications and radial
velocities are measured for objects classified as stars.
We now describe these three measurement procedures in turn.

Stellar velocity dispersions $\sigma_v$ of galaxies
are measured using a stellar template basis derived
from the ELODIE library \citep{Prugniel01},
covering the rest-frame
spectral range 4100\,\AA--6800\,\AA\@.
The high-resolution ELODIE spectra are degraded
to the binning scale and approximate
resolution of the co-added
BOSS spectra, and a PCA of the
library is performed.
The first 24 principal components
are used as a basis for fitting the
galaxy spectra.
The entire PCA basis is incrementally broadened from
0 to 850\,km\,s$^{-1}$ in units of 25\,km\,s$^{-1}$,
and the set of all broadened PCA components is cached
for the analysis of all galaxy spectra.
For each galaxy spectrum, the stellar PCA basis
is redshifted to match that galaxy's redshift.
At each trial broadening, the galaxy spectrum
is fit with an error-weighted
least-squares linear combination of the broadened
stellar PCA basis plus a quartic polynomial,
while masking the regions surrounding
common emission lines.
This marginalization over stellar-population effects
at each trial $\sigma_v$ value serves to
absorb some of the systematic errors
of ``template mismatch'' into the
statistical velocity dispersion error.
The $\chi^2$ goodness-of-fit statistic
for the best model
is tabulated for each broadening step, to
define a $\chi^2 (\sigma_v)$ curve.
The minimum-$\chi^2$ velocity-dispersion value
(with sub-grid localization) is reported as the
measured value \texttt{VDISP}, and the error on this measurement
\texttt{VDISP\_ERR} is estimated from the curvature of the
$\chi^2$ function at the position of the minimum.
Note that this analysis is highly analogous to the
redshift measurement procedure described
in \S\ref{subsec:zmeasure}.  This velocity dispersion
measurement algorithm is unchanged from SDSS-I/II\@.

Within the BOSS data set, the S/N per pixel in galaxy spectra is often
below the threshold commonly adopted as minimally sufficient for
accurate point estimation of the stellar
velocity dispersion.  
However, it has been shown by \citet{Shu12}
that unbiased measurements of the
\textit{distribution} of velocity dispersions within
a large sample of galaxies can be made even when the
individual spectra are of low S/N, by means
of a hierarchical analysis that marginalizes statistically
over the likelihoods of all possible
velocity-dispersion values for each galaxy.
To enable such analyses, we also compute
and report the velocity-dispersion likelihood function for each galaxy
in the vector-valued column \texttt{VDISP\_LNL}\@.  This is
defined by $- \chi^2(\sigma_v) / 2$ for velocity dispersions $\sigma_v$
from 0 to 850\,km\,s$^{-1}$ in steps of 25\,km\,s$^{-1}$.
The baseline and overall $\chi^2$ computation
method are the same as described above for the
measurement of the \texttt{VDISP} point estimators.
However, the \texttt{VDISP\_LNL} calculation employs only the first
five stellar PCA template basis spectra,
and also marginalizes over galaxy redshift uncertainties.
An additional difference is that while the \texttt{VDISP}
computations are done only for objects with \texttt{CLASS} of galaxy
(for consistency with the SDSS-I/II practice), the \texttt{VDISP\_LNL}
calculations are done only for objects with \texttt{CLASS\_NOQSO} of galaxy
(for consistency with BOSS practice).

Emission-line parameters for the 31 transitions listed
in Table~\ref{table:emlines} are computed for all spectra
for which those lines fall into the observed BOSS wavelength range.
Each line is modeled as a Gaussian, and the amplitudes,
centroids, and widths of all lines are optimized non-linearly
to obtain a minimum-$\chi^2$ fit to the data.
The background continuum spectrum is taken from the best-fit
velocity-dispersion model (for galaxies), from the
best-fit redshift-pipeline model (for stars),
and from a linear fit to the sidebands of each line
(for quasars and for ranges of the galaxy spectra that
extend beyond the coverage of the ELODIE-based
velocity-dispersion templates.)
All lines are constrained to have the same redshift within the
fit, with the exception of Ly$\alpha$, which is allowed to
fit at a different redshift to account for the
asymmetric effects of Ly$\alpha$ forest absorption.
In addition, groups of lines are constrained
to have the same line-width
as noted in Table~\ref{table:emlines},
so as to allow robust fits to the
strengths of low-S/N emission lines.
Hence, the reported line-widths are
effectively a strength-weighted average over the group.
An initial guess for the line redshifts is taken from the
best-fit pipeline redshift.  Emission-line redshifts are
allowed to depart arbitrarily from this value, but in practice
are well-constrained in cases with significant emission
in any lines.  96\% of the quasars with significant C\textsc{iv} emission
have line fits within 6000\,km\,s$^{-1}$ of the template redshift,
and 96\% of the galaxies with significant [O\textsc{ii}] emission have fits
within 100\,km\,s$^{-1}$.
Line fluxes, line widths, line redshifts, estimated
continuum levels, and observed-frame equivalent widths are reported
by the line fitting code, along with associated errors.
In the SDSS-I/II implementation of the \texttt{idlspec2d}
emission-line measurement code, equivalent widths
were measured relative to the estimated
continuum spectrum at line center, while for BOSS DR9 this has
been changed to use a continuum level estimated from
the sidebands of the line.

Based on the results of the line-fitting code,
galaxy spectra with emission in all four of the lines
H$\beta$, [O\textsc{iii}]~5007, H$\alpha$, and [N\textsc{ii}]~6583
detected at 3-sigma or greater
are sub-classified into \texttt{AGN}, \texttt{STARFORMING}, and
\texttt{STARBURST} according to the following rules.
First, galaxies are sub-classified as \texttt{AGN} if
\begin{equation}
\log_{10}([O\textsc{iii}]/H\beta) > 1.2 \log_{10}([N\textsc{ii}]/H\alpha) + 0.22
\end{equation}
\citep{Baldwin81}.
For galaxies falling on the other side of this cut,
sub-classification is made based on
the equivalent width of H$\alpha$:
\texttt{STARFORMING} if less than 50\,\AA,
and \texttt{STARBURST} if greater.
Galaxies and quasars may be given an additional sub-classification
as \texttt{BROADLINE} if they have line widths in excess
of 200\,km\,s$^{-1}$, with line-width measurement significance
of at least 5-sigma, and line-flux measurement
significance of at least 10-sigma.

\begin{table}[t]
\begin{center}
\caption{\label{table:emlines}
Emission lines measured by the BOSS pipeline}
\begin{tabular}{llll}
\hline \hline
Line & Line & Redshift & Width  \\
Wavelength\tablenotemark{1} & Name & Group\tablenotemark{2} & Group\tablenotemark{3} \\
\hline
1215.67  & Ly$\alpha$            & Ly$\alpha$ & Ly$\alpha$ \\
1240.81  & N\textsc{v} 1240      & emission   & N\textsc{v} \\
1549.48  & C\textsc{iv} 1549     & emission   & emission \\
1640.42  & He\textsc{ii} 1640    & emission   & emission \\
1908.734 & C\textsc{iii}] 1908   & emission   & emission  \\
2799.49  & Mg\textsc{ii} 2799    & emission   & emission  \\
3726.032 & [O\textsc{ii}] 3725   & emission   & emission \\
3728.815 & [O\textsc{ii}] 3727   & emission   & emission \\
3868.76  & [Ne\textsc{iii}] 3868 & emission   & emission \\
3889.049 & H$\epsilon$           & emission   & Balmer \\
3970.00  & [Ne\textsc{iii}] 3970 & emission   & emission \\
4101.734 & H$\delta$             & emission   & Balmer \\
4340.464 & H$\gamma$             & emission   & Balmer \\
4363.209 & [O\textsc{iii}] 4363  & emission   & emission \\
4685.68  & He\textsc{ii} 4685    & emission   & emission \\
4861.325 & H$\beta$              & emission   & Balmer \\
4958.911 & [O\textsc{iii}] 4959\tablenotemark{4}  & emission   & emission \\
5006.843 & [O\textsc{iii}] 5007\tablenotemark{4}  & emission   & emission \\
5411.52  & He\textsc{ii} 5411    & emission   & emission \\
5577.339 & [O\textsc{i}] 5577    & emission   & emission \\
5754.59  & [N\textsc{ii}] 5755   & emission   & emission \\
5875.68  & He\textsc{i} 5876     & emission   & emission \\
6300.304 & [O\textsc{i}] 6300    & emission   & emission \\
6312.06  & [S\textsc{iii}] 6312  & emission   & emission \\
6363.776 & [O\textsc{i}] 6363    & emission   & emission \\
6548.05  & [N\textsc{ii}] 6548\tablenotemark{5}   & emission   & emission \\
6562.801 & H$\alpha$             & emission   & Balmer \\
6583.45  & [N\textsc{ii}] 6583\tablenotemark{5}   & emission   & emission \\
6716.44  & [S\textsc{ii}] 6716   & emission   & emission \\
6730.82  & [S\textsc{ii}] 6730   & emission   & emission \\
7135.790 & [Ar\textsc{iii}] 7135 & emission   & emission \\
\hline
\end{tabular}
\tablenotetext{1}{Wavelengths are quoted in air for optical transitions
and in vacuum for UV transitions below 2000\,\AA.}
\tablenotetext{2}{Emission lines of a common ``redshift group'' are constrained
to have the same redshift in the line fitting procedure.}
\tablenotetext{3}{Emission lines of a common ``width group'' are constrained
to have the same intrinsic velocity width in the line fitting procedure.}
\tablenotetext{4}{[O\textsc{iii}] 5007 and [O\textsc{iii}] 4959
are constrained to have a 3:1 line-flux ratio.}
\tablenotetext{5}{[N\textsc{ii}] 6583 and [N\textsc{ii}] 6548
are constrained to have a 3:1 line-flux ratio.}
\end{center}
\end{table}

For spectra classified as stars, an additional fitting to the ELODIE
stellar library \citep{Prugniel01} is performed.
The ELODIE library contains 709 stars spanning spectral types O to M,
luminosity classes V to I, and metallicities [Fe/H] from $-3.0$ to $+0.8$.
The observed resolution was 42,000 over the wavelength range 4100 to 6800\,\AA\@.
Our fitting makes use of the release of this library at resolution 10,000
that was calibrated to 0.5\% in narrow-band spectrophotometric
precision and 2.5\% in broad-band precision.
This library was trimmed from 709 to 610 stars that are not binary
or triple stars.
The ELODIE spectra are convolved with Gaussian functions
to match the resolution of the BOSS spectra.
A later release of this library (ELODIE 3.1) was not used
due to extensive masking of regions near sky emission that compromises
its use for measuring radial velocities to high precision \citep{Prugniel07}.
Each BOSS spectrum classified as a star is re-fit to all spectra
in this trimmed ELODIE library with the identical redshift-fitting code
used to determine the primary redshift (\S\ref{subsec:zmeasure}).
These fits are limited to the 4100--6800\,\AA\ wavelength range,
include 3 polynomial terms, and span velocities from $-1000$ to $+1000$\,km\,s$^{-1}$.
The physical parameters of the best-fit ELODIE template are
included in the pipeline outputs
(\texttt{ELODIE\_TEFF}, \texttt{ELODIE\_LOGG}, \texttt{ELODIE\_FEH}),
along with the redshift (\texttt{ELODIE\_Z}),
statistical error of the redshift (\texttt{ELODIE\_Z\_ERR})
and reduced $\chi^2$ of that fit (\texttt{ELODIE\_RCHI2}).
An estimate of the template-mismatch effects on the redshift is
provided as the standard deviation in redshift among the best
12 ELODIE template fits (\texttt{ELODIE\_Z\_MODELERR})\@.

The BOSS pipeline also computes and reports
median spectroscopic signal-to-noise ratios per 69\,km\,s$^{-1}$ pixel
(\texttt{SN\_MEDIAN})
over the five SDSS broadband wavelength ranges ($ugriz$, \citealt{Fukugita96}),
along with the synthetic broadband fluxes predicted by the
spectrum (\texttt{SPECTROFLUX})
and the best-fit template model to the spectrum (\texttt{SPECTROSYNFLUX})\@.

As described in \citet{Nine12}, DR9 also
includes catalogs of alternative
parameter measurements for BOSS galaxies,
which are documented in other publications.
\citet{Chen12} describe PCA-based stellar-population
parameter measurements and velocity-dispersion estimates.
\citet{Thomas12} have measured
stellar velocity dispersions using the pPXF software
of \citet{Cappellari04} and emission-line properties
using the GANDALF software of \citet{Sarzi06}.
Finally, \citet{Maraston12} have derived photometric
stellar-mass estimates for BOSS galaxies.
All these measurements are distributed with DR9,
but are separate from the core \texttt{idlspec2d}
pipeline system described here.

\subsection{Output files}
\label{subsec:outfiles}

The BOSS \texttt{idlspec2d} redshift pipeline generates output files
for each plate, along with summary files to aggregate
photometric and spectroscopic parameters across the entire BOSS survey
data set.
These files are listed in Table~\ref{table:files};
together they contain all the parameters described in this paper.
Access to these files on the SDSS-III Science Archive Server (SAS),
as well as full data-model documentation of their formats and contents,
can be obtained through the SDSS-III DR9 website.
The \texttt{spAll} summary file from the BOSS
pipeline is analogous but not identical in form and
content to the \texttt{specObj} file loaded by the
SDSS-III Catalog Archive Server (CAS),
which contains both SDSS-I/II and BOSS data.

\begin{table}[t]
\begin{center}
\caption{\label{table:files} DR9 Redshift and classification pipeline output files\tablenotemark{1}}
\begin{tabular}{ll}
\hline \hline
\texttt{spZbest-pppp-mmmmm.fits} & Best-fit redshift \& class param.s \\
\texttt{spZall-pppp-mmmmm.fits} & Parameters for all fits \\
\texttt{spZLine-pppp-mmmmm.fits} & Emission-line parameters \\
\texttt{spAll-v5\_4\_45.fits} & Summary param.s for all spectra \\
\texttt{spAllLine-v5\_4\_45.fits} & Line fit param.s for all spectra \\
\hline
\end{tabular}
\tablenotetext{1}{The strings \texttt{pppp} and \texttt{mmmmm} represent the 4-digit \texttt{PLATE}
and 5-digit \texttt{MJD} identifiers for files that are created on a plate-by-plate basis.
The string \texttt{v5\_4\_45} denotes the frozen version of the \texttt{idlspec2d} software
used for the processing of the DR9 spectroscopic data sample.
Full documentation of these and other
pipeline output files are found at
\url{http://www.sdss3.org/dr9/} \\}
\end{center}
\end{table}

Approximately 8\% of BOSS spectra are repeat observations
of previously observed targets, due both to re-observations of
entire plates and to re-targeting of a number of
objects on more than one plate (see \citealt{Dawson12}).
Of particular note within the summary files, the best spectroscopic
observation of each object (defined by a 2$^{\prime\prime}$ positional match)
in the survey is defined according to the following rules:
\begin{enumerate}
\item Prefer spectra with positive median S/N per
spectroscopic pixel
within the $r$-band wavelength range
over other observations.
\item Prefer spectra with \texttt{ZWARNING}~$ = 0$ over other
spectra (or \texttt{ZWARNING\_NOQSO}~$ = 0$ for galaxy-sample targets.)
\item Prefer spectra with higher median S/N per spectroscopic pixel
within the $r$-band wavelength range.
\end{enumerate}
The best observation for each object is designated by setting
the parameter \texttt{SPECPRIMARY} equal to 1 in
the \texttt{spAll} file, while setting it
equal to zero for all other spectroscopic observations of a given
object that may be present within the survey data set.


\section{Template Classes}
\label{sec:templates}

In order to compare and select among galaxy, quasar, and stellar
models objectively and with the highest statistical
significance, the BOSS pipeline requires
redshift and classification measurement templates that
span both the full space
of physical object types within the survey and
the full wavelength range of the spectrograph.
BOSS expands on SDSS-I/II in both regards,
and hence requires a new set of pipeline templates,
which we now describe.

\subsection{Galaxies}
\label{subsec:galtemp}

The \texttt{idlspec2d} galaxy redshifts for SDSS-I/II
were measured using templates generated from 480 galaxies observed
on SDSS plate 306, MJD 51690.\footnote{These spectra are tabulated
in the file \texttt{eigeninput\_gal.dat} within the \texttt{templates}
subdirectory of the \texttt{idlspec2d} product.}
Redshifts for this training set were
established by modeling each spectrum across
a range of trial redshifts as a linear combination of
(1) the leading two components of a PCA analysis of
10 velocity-standard stars in M67 observed by SDSS-I plate
321 on MJD 51612, (2) a set of common
optical emission lines modeled as narrow Gaussian profiles, and (3) a low-order
polynomial.  The adopted redshift for each galaxy was taken
from the location of the minimum-$\chi^2$ value localized to sub-grid accuracy,
in the same manner described in \S\ref{subsec:zmeasure} above.
Using these redshifts, the training-sample spectra were transformed
to a common rest-frame baseline, and input to an iterative
PCA procedure that accounts for
measurement errors and missing data (e.g., \citealt{Tsalmantza12}
and references therein).
The leading four ``eigenspectra'' from this procedure
were taken to define the galaxy redshift template basis
for SDSS-I/II\@.
For the commissioning analysis of BOSS spectra, these same templates were
used for measuring galaxy redshifts, despite their lack of $z=0$
coverage redward of 9300\,\AA\ and their under-representation
of post-starburst galaxies (which appear with more
frequency in the BOSS CMASS sample than in SDSS-I/II)\@.

To generate a new redshift template set for use in automated analysis of
BOSS spectra, we select a set of BOSS galaxies with
redshifts over the interval $0.05 < z < 0.8$ that are
well-measured by the original SDSS templates.
To increase S/N and flatten the coverage of galaxy
parameter space before performing a PCA to generate the
template set, we bin together galaxies with similar 4000\,\AA\
break strengths ($D4000$, \citealt{Balogh99}) and redshifts, for the
purposes of stacking their spectra.
We use a $D4000$ range from 1.0 to 2.2 with a binning interval
of 0.2, and a redshift binning interval
of 0.05. In some $D4000$--redshift bins, we further subdivide
the galaxies into several H$\delta_A$ sub-bins.  The number of
sub-bins depends on the number of galaxies in each $D4000$--redshift bin:
if the total number is smaller than 600, we do not divide further
into H$\delta_A$ sub-bins; if the number is in the
range 600--1200, we divide into two sub-bins; and if there are greater
than 1200, we divide into three sub-bins.

We also select a set of ``post-starburst'' galaxies from
the BOSS galaxy sample,
defined by having either
\begin{equation}
D4000<1.3 ~~\mbox{and}~~ (H\delta_A + H\gamma_A)/2 > 7
\end{equation}
or
\begin{equation}
(H\delta_A + H\gamma_A)/2 > {\rm max}[-17.50 \times
D4000 + 29.25, ~3].
\end{equation}
This criterion leads to a sample of 
about 2400 post-starburst
galaxies, which we divide into five bins in redshift
with equal numbers of galaxies per bin.
We then stack the rest-frame spectra of all galaxies
in each bin, to generate a set of high-S/N stacked spectra
across the range of parameters indicated.

\begin{figure}[t]
\epsscale{1.15}
\plotone{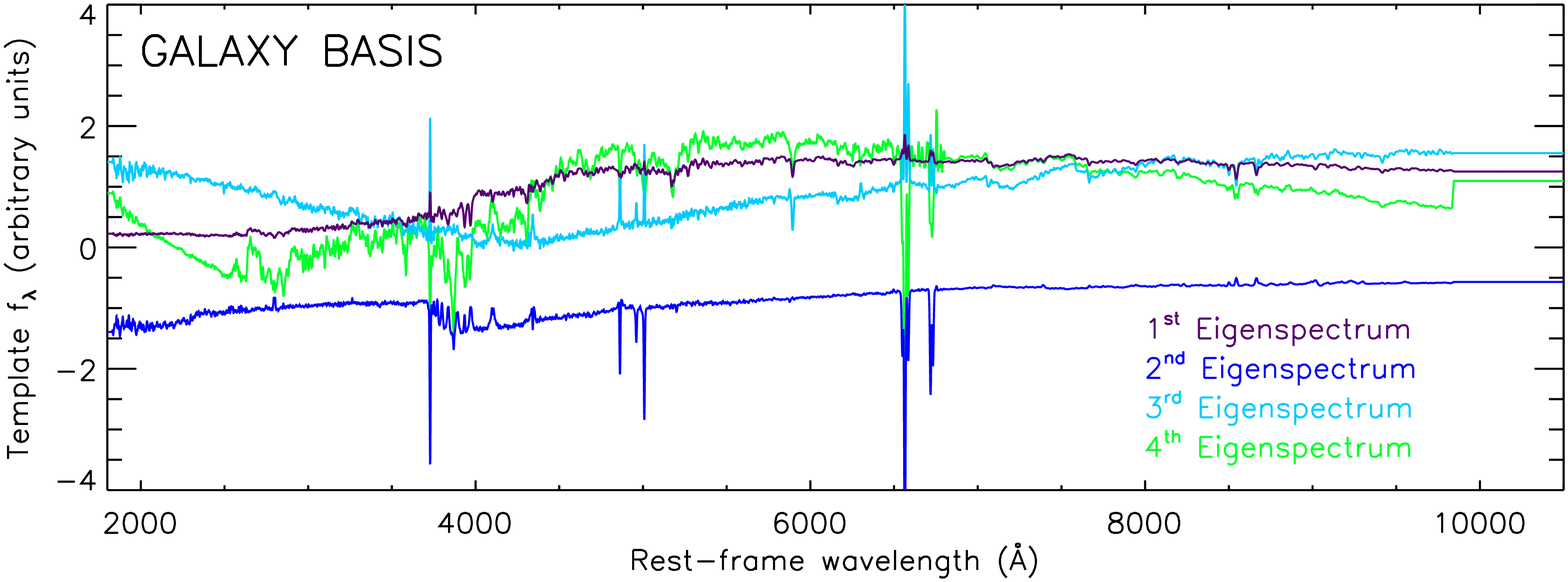}
\plotone{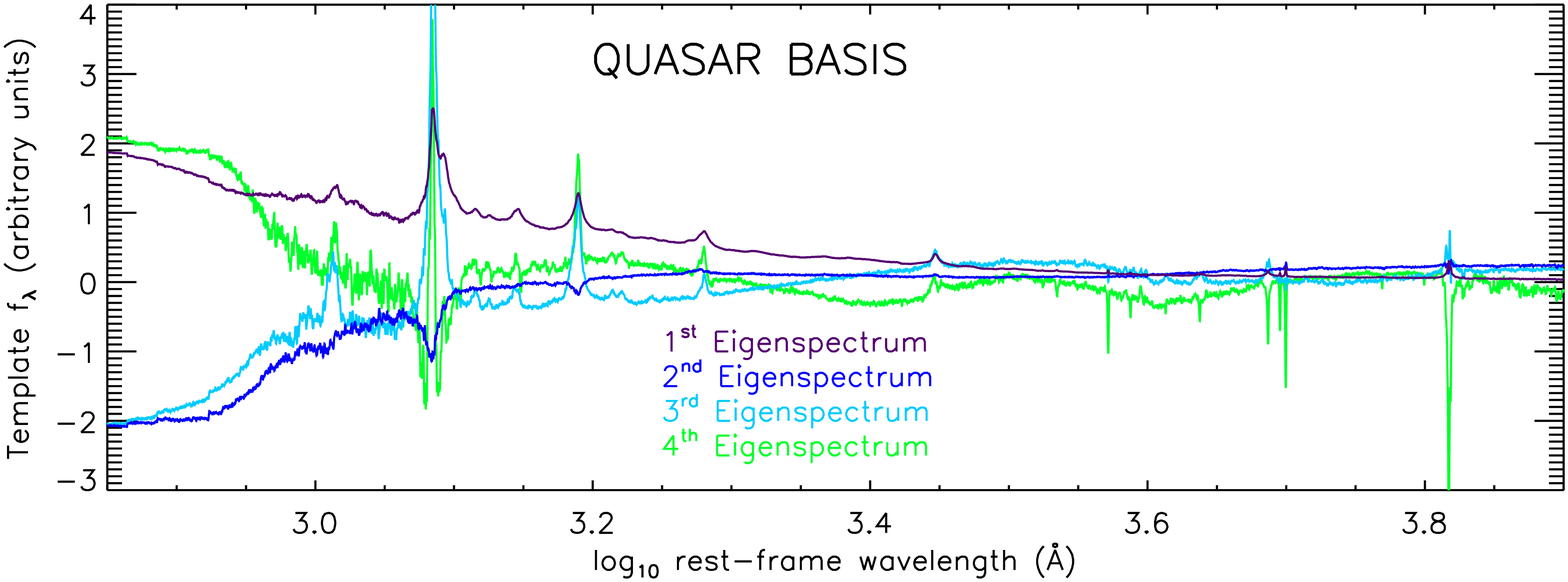}
\plotone{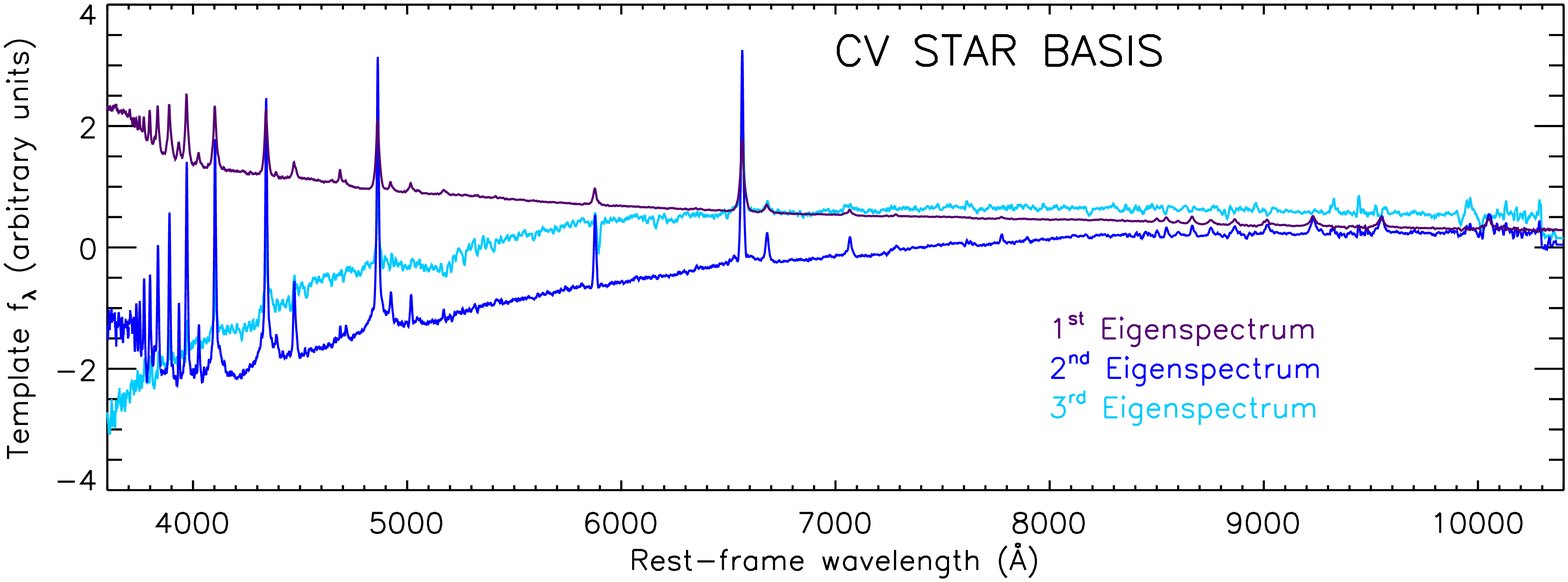}
\caption{\label{fig:templates}
BOSS redshift and classification template basis sets for
galaxies (\textit{top}), quasars (\textit{middle}), and CV stars (\textit{bottom}).
}
\end{figure}

Once all these stacked spectra are in hand,
we fit stellar continuum models to them
\citep{Brinchmann04,Tremonti04} using 
simple stellar population (SSP) models.
Our SSP models are taken from \citet{Maraston09}
and \citet{Maraston11}, and are based
on a combination of theoretical
and observational stellar library
data from \citet{Rodriguez05},
\citet{Sanchez06}, and
\citet{Gustafsson08}.
In the rest-frame wavelength 
range 1900--9900\AA,
we patch the stacked spectra with the fitted continuum models for pixels with S/N 
smaller than 10, pixels where the difference between models and stacks
is larger than 30\%, and pixels where there are no observations. We 
also add H$\alpha$, [N\textsc{ii}], and [S\textsc{ii}]
emission lines for cases where these lines
fall outside the range of observed wavelengths used to generate 
the stacked spectra.  This is accomplished by selecting galaxies with similar $D4000$,
H$\delta_A$, and dust extinction as the galaxies used to make the stacks,
and computing the median values of
$\sigma_{\rm H\alpha}/\sigma_{\rm H\beta}$,
$f_{\rm H\alpha}/f_{\rm H\beta}$,
$\sigma_{\rm [N\textsc{ii}]}/\sigma_{\rm [O\textsc{iii}]}$,
$f_{\rm [N\textsc{ii}]}/f_{\rm [O\textsc{iii}]}$,
$\sigma_{\rm [S\textsc{ii}]}/\sigma_{\rm [O\textsc{iii}]}$, and
$f_{\rm [S\textsc{ii}]}/f_{\rm [O\textsc{iii}]}$
for these comparison samples
(here, $\sigma$ is Gaussian line dispersion and $f$ is line flux). 
By multiplying these ratios with the appropriate
line width or flux of H$\beta$ or [O\textsc{iii}] from the 
stacks, we predict the line widths and fluxes for
H$\alpha$, [N\textsc{ii}] and [S\textsc{ii}] to be added to the stacked spectra,
which we do using a Gaussian model for each line.

At the end of this process, we have 160 stacked and patched
spectra.  We augment these data
with a sample of 28 type-II quasars
(e.g., \citealt{Zakamska03,Reyes08}) identified
within the BOSS spectroscopic data set (see the discussion in \S\ref{sec:issues}).
This full set of spectra is then used as input to
the rest-frame spectrum PCA algorithm
to generate the four-component BOSS galaxy redshift template basis,
which is shown in the top panel of Figure~\ref{fig:templates}.

\subsection{Quasars}
\label{subsec:qsotemp}

\begin{figure}[t]
\epsscale{1.2}
\plotone{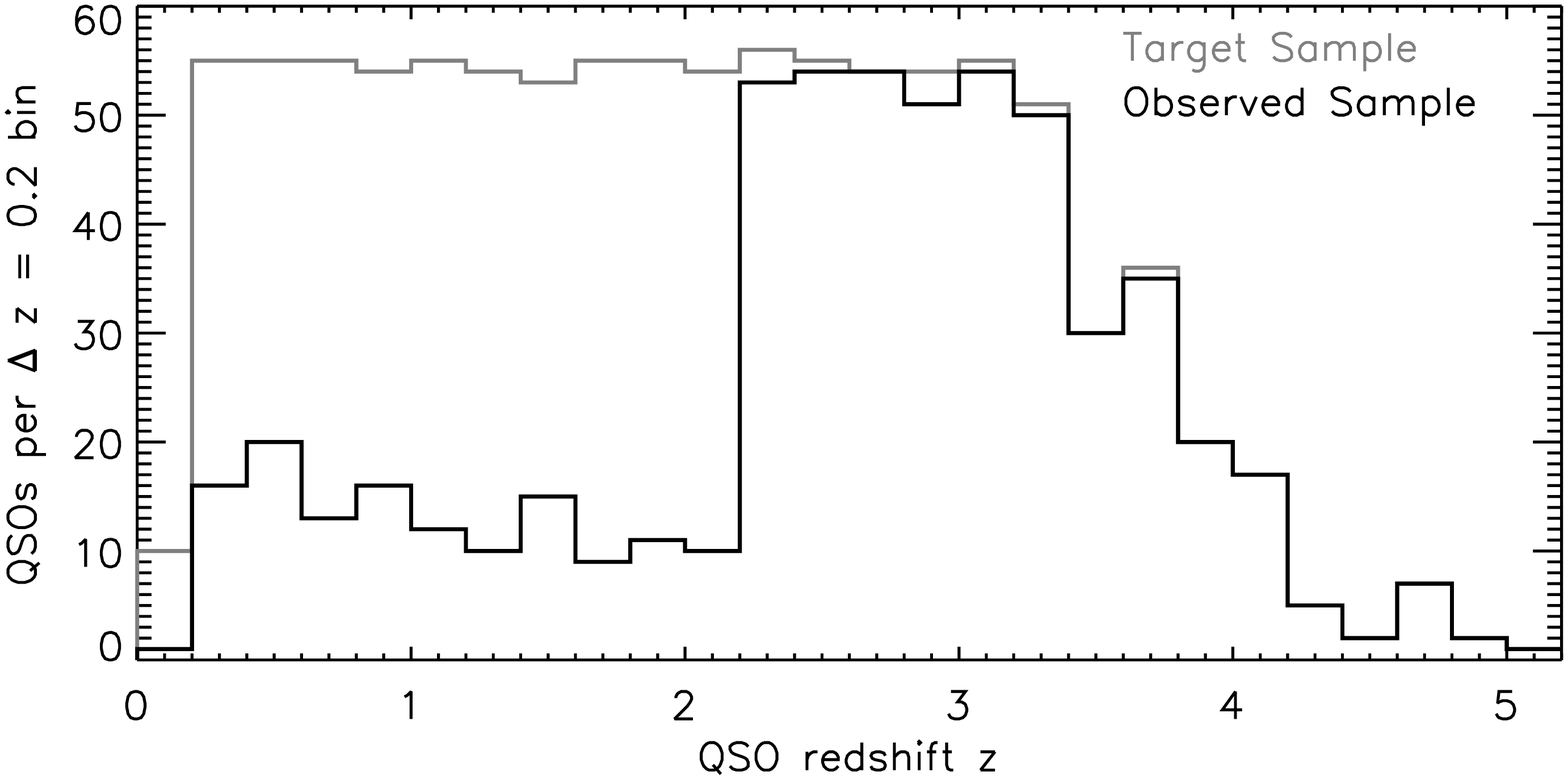}
\caption{\label{fig:qsohist}
Redshift distribution of 1000 targeted (gray) and 571 observed (black)
quasar training spectra.  Spectra from the observed distribution
are used to construct the PCA-based quasar redshift templates
used for automated classification and redshift
measurement in BOSS DR9 and shown in the middle
panel of Figure~\ref{fig:templates}.}
\end{figure}

Quasar redshift templates are generated from a training
sample of targets selected
from the SDSS DR5 quasar catalog \citep{Schneider07} and targeted for re-observation
with the BOSS spectrographs.  The targets were chosen from
the catalog at random, while
enforcing as uniform a distribution as possible in redshift.
As of 2011 June 10, 571 objects from this sample
had been observed by BOSS\@.  Removal of three
spectra for localized cosmetic defects gives a training
sample of 568 BOSS quasars.  The distribution in redshift of the targeted
sample and the observed sample is shown in Figure~\ref{fig:qsohist}.
The observed sample is weighted
more heavily above redshift $z = 2.2$, in accordance with
overlapping BOSS quasar sample priorities.  We keep this
weighting in the training set, since we want our redshifting
performance to be particularly well tuned for the redshift
range of interest to the BOSS Ly$\alpha$ forest program.

Using the redshifts given by \citet{Schneider07}, we shift
these training spectra to their rest frames and perform a PCA
of the sample, with iterative replacement to fill in
missing data.  The top four principal components are retained
and used as the linear basis set for our automated redshift
and classification measurements, and are shown in the middle
panel of Figure~\ref{fig:templates}.

We do not employ the redshift estimates of \citet{Hewett10}
for the quasar-template
training sample because the current BOSS pipeline
is not configured to incorporate the absolute-magnitude
information that would be
necessary to take advantage of the increased precision
afforded by these redshifts.  Future BOSS pipeline versions
may incorporate the \citet{Hewett10} approach.
We note that the primary criterion for BOSS spectroscopic
pipeline performance on quasar targets is to minimize
catastrophic redshift failures.  Several detailed approaches
to maximizing quasar redshift precision are
being investigated within the BOSS quasar science working
group, but all of these rely on having essentially correct
initial quasar redshifts from the \texttt{idlspec2d}
pipeline and/or visual inspection procedures \citep{Paris12}.

\subsection{Stars}
\label{subsec:startemp}

Although stellar science is not a primary goal of BOSS,
the redshift pipeline must successfully
flag stars from within the galaxy and quasar target samples of the survey. 
There is currently no comprehensive library of observed stellar spectra
covering the full usable wavelength range of the BOSS spectrograph
and the full H-R diagram.
To assemble a set of stellar templates suitable to BOSS spectrum
classification, we use a hybrid approach that extends data
from the Indo-US observational stellar spectrum library
\citep{Valdes04}, selected to provide
uniform coverage of the space of stellar atmosphere
parameters $T_{\mathrm{eff}}$, $\log g$, and [Fe/H]) using
theoretical atmosphere models computed using
the MARCS \citep[][for cool stars]{Gustafsson08},
ATLAS \citep[][for intermediate stars]{Kurucz05}, and
CMFGEN \citep[][for hot stars]{Hillier98} codes,
obtained via the curated POLLUX spectrum database \citep{Palacios10}.

\subsubsection{Template spectrum creation}

We start with the full database of 1273 Indo-US stellar spectra,
which have a resolution of approximately 1\,\AA, a reduced
pixel scale of 0.4\,\AA, spectral
coverage over 3400\,$<\lambda<$\,9500,
and good flux calibration for most stellar types.
The original radial-velocity zeropoints for the
library were established
either from literature or from velocity-standard
cross-correlations.  Since classification is the primary function
of these spectra within the BOSS pipeline,
we do not attempt any further refinement of these
velocity zeropoints.

We initialize a ``bad pixel'' mask for each Indo-US spectrum based upon the
zero-value Indo-US pixel mask convention.
Furthermore, we define the following telluric absorption bands,
and mask all pixels within them:
6850\,\AA--6950\,\AA,
7150\,\AA--7350\,\AA,
7560\,\AA--7720\,\AA,
8105\,\AA--8240\,\AA,
$>$8900\,\AA\@.
We then select the subset of spectra that
meet the conditions of (1) wavelength coverage from
at least 3500\,\AA\ to 8900\,\AA, (2) good data over at least 75\% of their pixels,
(3) flux calibration with a non-flat (i.e., stellar) standard, and (4) no single
gap within the spectrum
larger than 200\,\AA\ (the largest adopted telluric band width).
These cuts result in a sample of 879 spectra covering spectral types from
O6 to M8, but exclude carbon stars (which are fluxed with a flat SED
in the Indo-US library).

We then take the 1040 model atmosphere spectra from the POLLUX database
ranging in temperature from 3000\,K to 49000\,K, convolved and binned to 
the resolution and sampling of the Indo-US spectra.  For each Indo-US
spectrum in our subset, we loop over all model atmospheres and determine the
multiplicative scaling of the model that minimizes the sum of squared data-minus-model residuals over
non-masked pixels.  We adopt the model spectrum that gives the overall
minimum sum of squared residuals as being the ``best fit'' for a
particular data spectrum.

The ``best fit'' model spectrum for each data spectrum is used to extend the
data wavelength coverage and interpolate over the data gaps as follows.
We define a running window of $\pm 400$ pixels ($\pm 160$\,\AA) about an output
pixel of interest, and determine the multiplicative scale and tilt to apply to
the model over that window in order to give the best (least squares) fit to the
non-masked data pixels over that same range.  The scaled and tilted model is evaluated
at the central pixel to define the new, locally scaled model spectrum,
and the process is repeated over the entire spectrum by sliding the window.
For pixels centered outside the outermost pixel of data coverage on the red and blue ends,
the scale and tilt at the outermost data-covered pixel are used.
The data and ``sliding-scaled'' model spectra are combined into a single
output spectrum by assigning 100\% model in pixels where the
data have no coverage, defining a 100-pixel (40\,\AA) transition
region on either side of data gaps where the output spectrum is a weighted combination
of the data and the sliding-scaled model, and varying the weight linearly from
0\% model $+$ 100\% data to 100\% model $+$ 0\% data over the transition region.
Finally, we convolve and bin these output spectra down to the
typical resolution (about 3\,\AA\ FWHM) and reduced-spectrum sampling
($\Delta \log_{10} \lambda = 0.0001$ per pixel)
of the BOSS data, also transforming
from air to vacuum wavelengths to match the BOSS spectrum convention.

\subsubsection{Archetype subset selection}

From these 879 patched and extended stellar spectra, our goal is to select a representative subset
of ``archetypes'' that provide sufficient coverage of stellar parameter space
to perform automated spectroscopic star--galaxy and star--quasar separation, while not attempting
overly detailed stellar analysis that is beyond the scope of the BOSS science mission
(cf.\ \citealt{Lee08}).

We first visually inspect the template database and remove a single spectrum
with noticeable data quality issues in an unmasked data region
(Indo-US ID\#33111, 5450\,\AA\ $< \lambda < $ 6000\,\AA).  We also select the
12 template spectra that have significant emission lines, and retain
each of them for our final archetype set.  This leaves 866 spectra from which to select
the remainder of our archetype sample.

To select a subset of archetypes from the remaining set of
templates, we wish to make use of a measure of the degree of
similarity or difference between any two spectra.
We first restrict our attention to
the wavelength range 3400\,\AA--11000\,\AA, corresponding to
$N_{\mathrm{pix}} = 5099$ pixels at the
processed 69\,km\,s$^{-1}$ BOSS spectrum pixel scale.
We then re-normalize all the template spectra
to satisfy
\begin{equation}
\sum_{i=1}^{N_{\mathrm{pix}}} f_i^2 = N_{\mathrm{pix}} ~,
\end{equation}
where $f_i$ is the flux density (in the $f_{\lambda}$ sense) in pixel $i$.
We define a statistic $s^2$
measuring the quality of spectrum $f^{\prime}$, scaled by a factor $a$,
as a model for spectrum $f$:
\begin{equation}
s^2 = \sum_{i=1}^{N_{\mathrm{pix}}} (f_i - a f^{\prime}_i)^2 ~.
\end{equation}
With our normalization convention, the best-fit (minimum-$s^2$) scaling
is simply given by
\begin{equation}
a_{\mathrm{best}} = N^{-1}_{\mathrm{pix}} \sum_{i=1}^{N_{\mathrm{pix}}} f_i f^{\prime}_i ~,
\end{equation}
and the value of $s^2$ at this best scaling is given by
\begin{equation}
s^2_{\mathrm{best}} = N_{\mathrm{pix}} (1 - a_{\mathrm{best}}^2) ~.
\end{equation}
Note that $a_{\mathrm{best}}$ and $s^2_{\mathrm{best}}$ are symmetric
under the interchange of $f$ and $f^{\prime}$: the amplitude
and fit quality
of one template to another does not depend upon which one is
taken as the ``data'' and which one as the ``model''.  Thus
$s^2_{\mathrm{best}}$ can be regarded as a measure of how different
two templates are from one another.

We compute the matrix of $s^2_{\mathrm{best}}$ between all pairs of
templates in our set, and determine our archetype list in an iterative
procedure.  We set a threshold of 7.5 for the maximum $s^2_{\mathrm{best}}$
allowable between two spectra in order for one spectrum to be an acceptable
representative for the other.
This threshold was selected heuristically to tune
the size and diversity of the final sample.
We then identify the single template spectrum within
the sample that has the most $s^2_{\mathrm{best}} < 7.5$ matches to the rest
of the sample.  This spectrum and all the spectra that it matches are
removed from further
consideration, and the process is iterated until all spectra have been
accounted for in this manner.  For our chosen threshold, this process
identifies 105 archetypes out of 866 analyzed templates.  When added to the
12 emission-line templates, this yields a set of 117 stellar
templates for our automated spectrum classification algorithm.
Spectra fit by these templates are tagged with the stellar
subclass listed in the Indo-US database, along with the library
identification number of the archetype spectrum.

\subsubsection{Special stellar subclasses}

Several subclasses of star appear with some frequency in the BOSS
target sample, but are not represented in the (flux-calibrated) Indo-US library.
For these subclasses, representative training samples
from within the BOSS data set are identified based upon classification
using SDSS-I/II stellar templates.  New templates are
derived by averaging the spectra of these training sets
within a PCA framework.  The six subclasses and the number of training spectra
for each of them are: (1) 47 carbon stars; (2) 50 ``hotter'' white dwarfs
with $u - g < 0.3$; (3) 50 ``cooler'' white dwarfs
with $u - g \ge 0.3$; (4) 19 calcium white dwarfs; (5) 31 magnetic
white dwarfs; and (6) 50 L dwarfs.  In addition, a sample of 18
cataclysmic variable stars (CVs) with prominent emission lines
is used to define a CV star eigenbasis of 3 PCA modes,
which is shown in the bottom panel of Figure~\ref{fig:templates}.
Because of the use of multiple eigenvectors rather than a single
average spectrum, CV stars are treated as an object class separate from
other stars in the automated classification analysis.

\section{Performance and Verification}
\label{sec:performance}

\begin{figure*}
\epsscale{1.17}
\plottwo{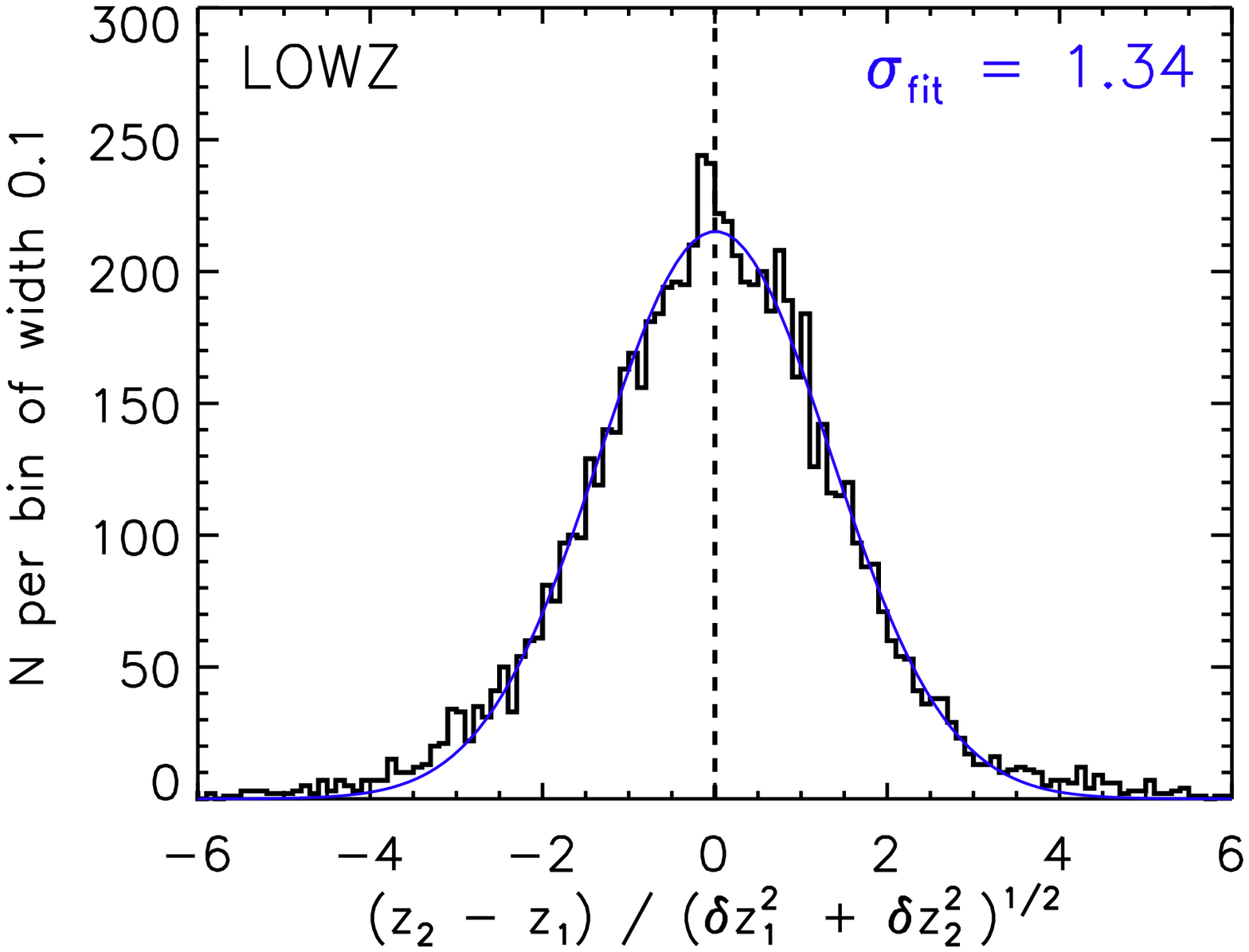}{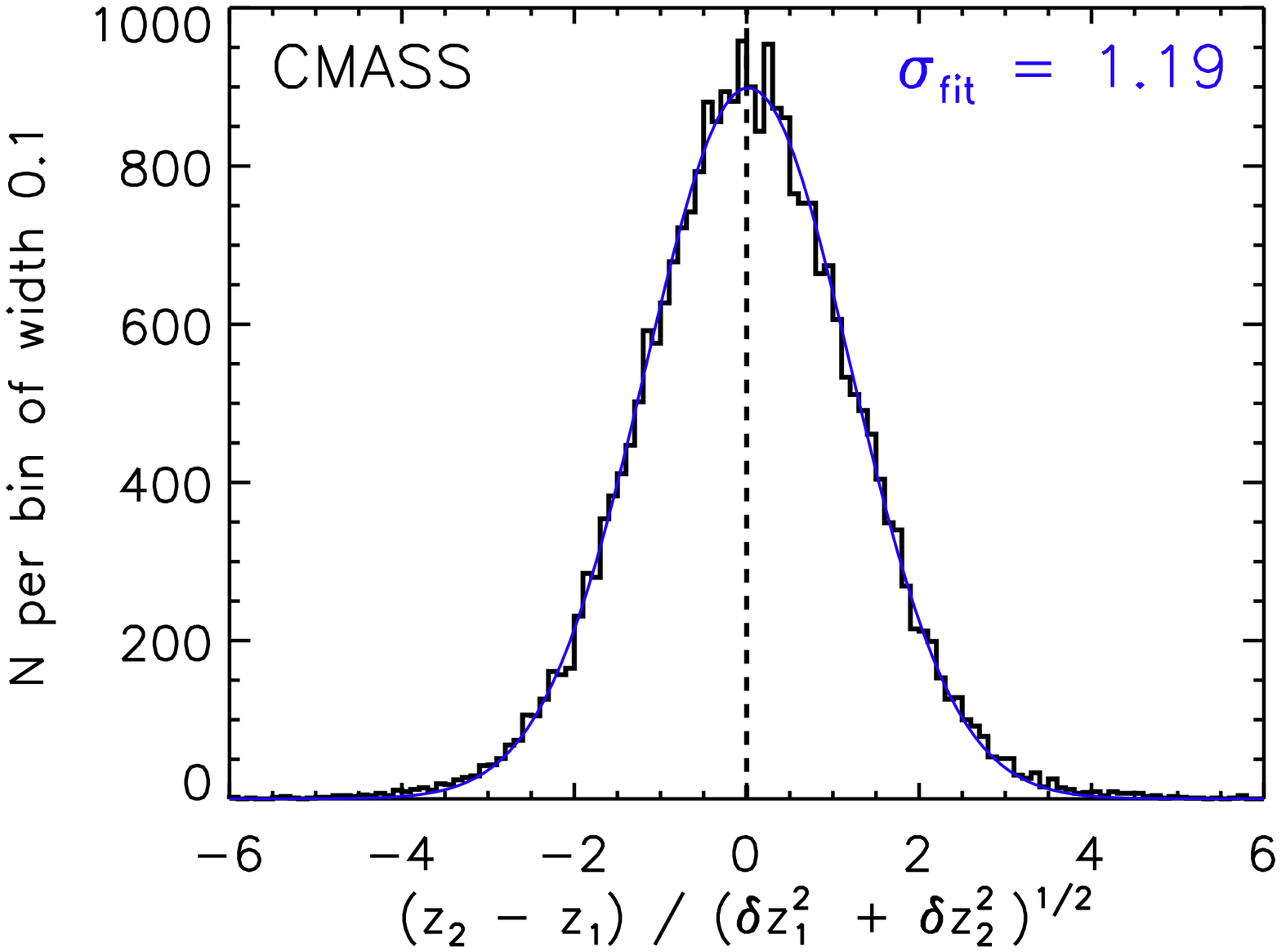}
\caption{\label{fig:z_err_hist}
Histograms of redshift differences
of LOWZ (left) and CMASS (right) galaxies that are observed
more than once, scaled
by the quadrature sum of statistical error estimates
in each epoch.  Over-plotted are the best-fit Gaussian models,
with a dispersion parameter of $\sigma=1.34$
for the LOWZ sample and $\sigma=1.19$ for the CMASS
sample.}
\end{figure*}

Table~\ref{table:dr9summary} provides a summary of the BOSS DR9
spectroscopic data set analyzed by the redshift and classification
pipeline described in this work, along with
a number of summary performance statistics that we now examine.
Additional checks on the \texttt{idlspec2d} pipeline performance
for galaxy targets in comparison with the \texttt{zcode} cross-correlation redshift
software described by \citet{Cannon06} are presented in \citet{Dawson12},
and additional discussion of pipeline quasar classification and redshift
performance is found in \citet{Paris12}.
The BOSS DR9 sample contains 831,000 spectra.
Of these, about 0.2\% are lost to
unplugged fibers and spectra falling along bad CCD
columns.  Approximately 92\% of the BOSS DR9 spectra
are of unique objects (as defined by a 2$^{\prime\prime}$ positional
match).  The remaining 8\% are repeat spectra
from overlapping plates or repeat observations
of the same plate.

\subsection{Galaxy redshift completeness and purity}
\label{subsec:galcomp}

\begin{figure*}
\epsscale{1.17}
\plottwo{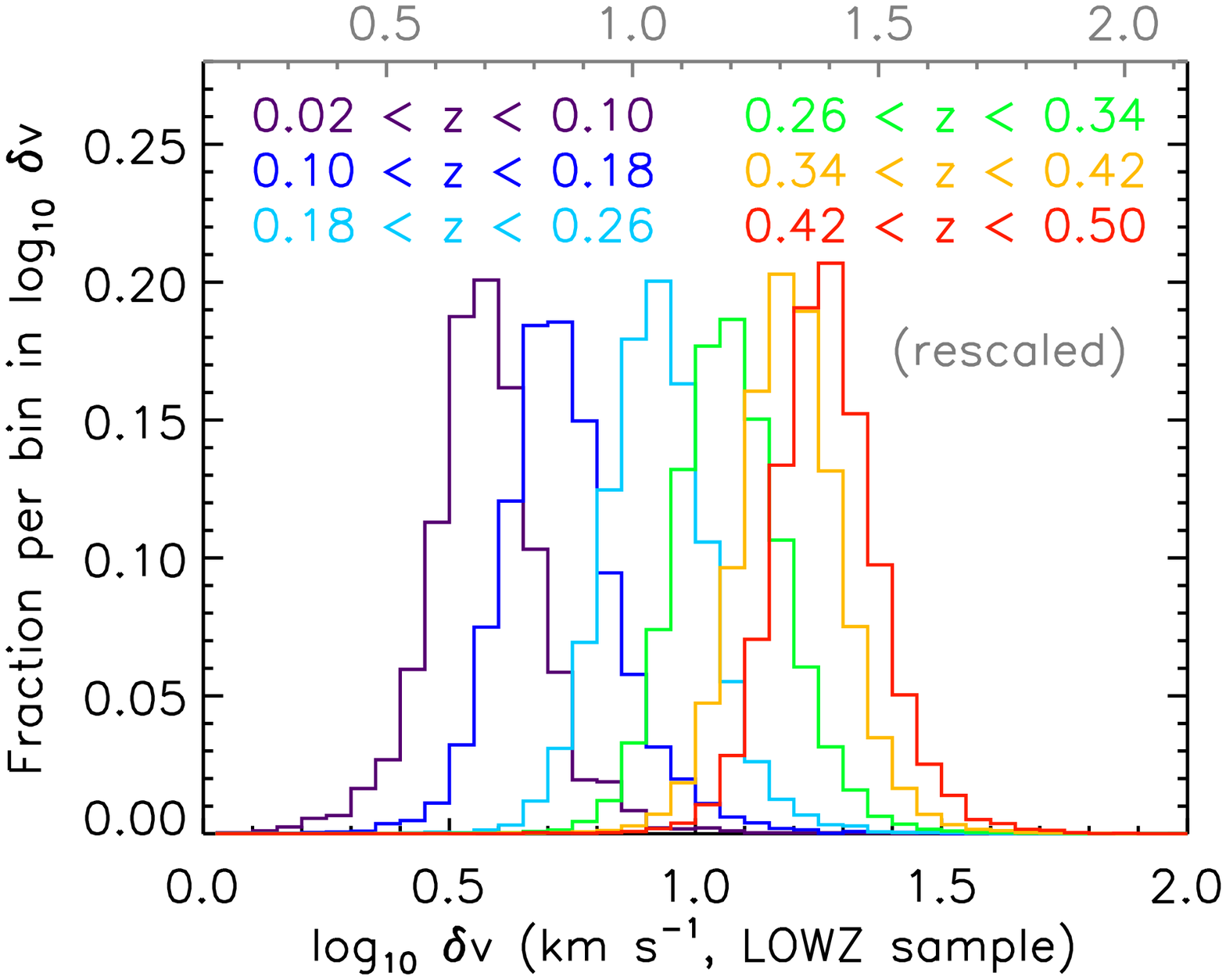}{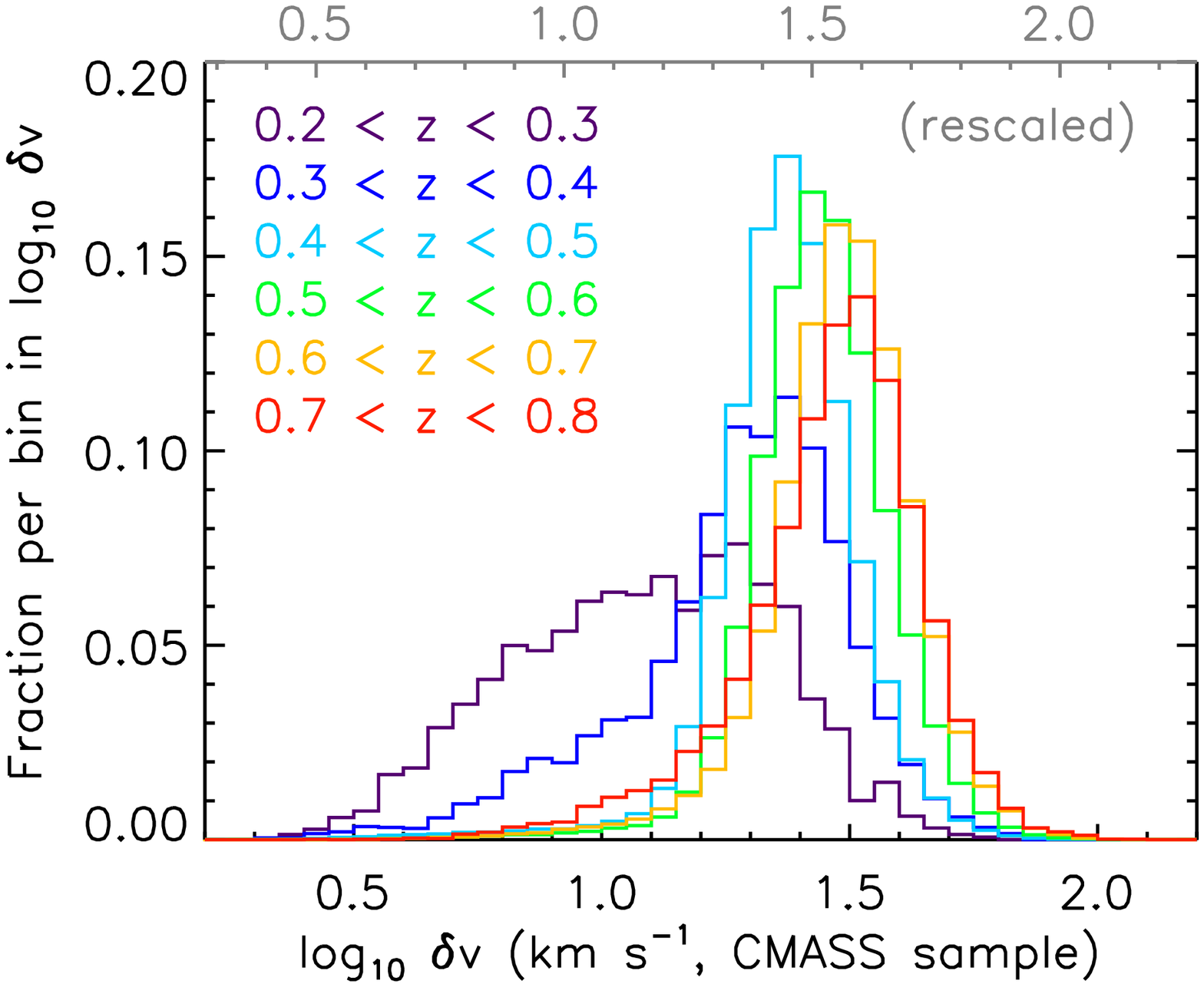}
\caption{\label{fig:v_err_hist}
Distribution of estimated \textit{statistical} galaxy redshift errors
for LOWZ (left) and CMASS (right) sample galaxies over multiple redshift ranges.
The horizontal axes at bottom indicate raw error estimates;
the gray horizontal axes at top indicate errors
rescaled by the factors illustrated
in Figure~\ref{fig:z_err_hist}.}
\end{figure*}

Using the \texttt{Z\_NOQSO} redshift measurement convention
as described in \S\ref{subsec:z_noqso}, we achieve an
automated completeness (i.e., \texttt{ZWARNING\_NOQSO == 0} rate)
of 98.7\% for the CMASS sample and
99.9\% for the LOWZ sample (from Table~\ref{table:dr9summary}).
Restricting further to objects that are spectroscopically
classified as galaxies (\texttt{CLASS\_NOQSO == "GALAXY"}),
we find combined targeting and measurement completeness
percentages of 95.4\% for CMASS and 99.2\% for LOWZ\@.
These percentages satisfy the BOSS science requirement
of at least 94\% overall galaxy redshift success.  For the CMASS sample, about 70\% of
the (small) survey inefficiency is due to targeting stars and
star--galaxy superpositions rather than
galaxies, and about 30\% arises from known redshift measurement failures.

To verify the completeness and quantify the purity of the automated
galaxy redshifting and classification, we make use of a ``truth table''
generated by the first author from the visual inspection of
4864 galaxy spectra taken on eight plates observed during
2010 March.\footnote{The
plates are: 3804 of MJD 55267; 3686, 3853, and 3855 of MJD 55268; and
3687, 3805, 3856, and 3860 of MJD 55269.}
We focus primarily on the CMASS sample, as this higher-redshift
(and thus lower S/N) sample poses the greatest challenge to the software.
Of the inspected spectra, 3666 are CMASS targets that are above the
fiber-magnitude threshold, not unplugged, and
not falling on bad CCD columns.  From among these 3666
galaxy-sample spectra, 3627 have confidently
measured pipeline redshifts and classifications, giving an automated
completeness of 98.9\%, consistent with the completeness of the
full DR9 CMASS sample from above.  Of this subset, 3500 are classified
as galaxies (as opposed to stars) by the pipeline, giving a 95.5\%
overall sample completeness including target-selection efficiency,
which is also consistent with the sample-wide value.

To quantify the purity of the CMASS spectroscopic redshift sample,
we first search for ``catastrophic'' impurities in the CMASS
redshift sample, defined as spectra for which the pipeline reports
a confident galaxy classification and redshift, but for which
the visual inspection yields a confident classification (of any class)
with a redshift that differs by greater than $\Delta z = 0.005$.
This search yields three such spectra out of 3500: two are
definite galaxy--M-star superpositions, and the other is a
possible galaxy-galaxy superposition (for which the pipeline redshift
is more convincing in retrospect than the inspection redshift).
We next check for less clearly defined impurities,
defined as spectra for which the pipeline reports
a confident galaxy classification and redshift, but for which
the visual inspection does not yield a confident result.
This search identifies 10 such spectra, six of which are
plausible pipeline redshifts with subjective
visual judgments of low S/N, and the remaining four of which
are due to artifacts associated with spectrum combination
across the spectrograph dichroic at 6000\,\AA\
(see Item~\ref{item:crosstalk} in \S\ref{sec:issues}).
Taking the three superposition spectra
and the four artifact spectra as genuine contaminants,
we find a CMASS sample impurity rate of about 0.2\%,
satisfying the 1\%  maximum catastrophic
redshift failure rate specified as the scientific requirement for BOSS\@.

To check for the possibility of recoverable incompleteness,
we examine CMASS spectra for which the visual inspections
yield a confident galaxy classification and redshift, but for which
the automated pipeline yields either no confident result
(i.e., \texttt{ZWARNING\_NOQSO > 0}),
or a classification as a star.  There are 26 such spectra,
which break down as follows: 11 low-S/N spectra for which the
pipeline's lack of confidence is statistically defensible;
5 definite or possible galaxy--galaxy superpositions; 4
definite or possible star--galaxy superpositions; 3 spectra
with artifacts; 2 broadline AGN mistaken for stars (but with
correct quasar-class redshifts that are
excluded by the \texttt{Z\_NOQSO} convention);
and 1 narrow-line AGN for which the pipeline confuses [O\textsc{iii}] 5007
and H$\alpha$.  Taking the 11 noisy but visually convincing redshifts
and the three AGN spectra to represent the recoverable sample,
we find an excess incompleteness of about 0.4\% relative to the maximum
attainable given the data.

To further assess the effects of star--galaxy superpositions
(for which the pipeline takes no special approach),
we search a set of 57910 CMASS spectra from 150 plates
for instances of a best-fit non-quasar class of GALAXY and a
next-best non-quasar class of STAR, and examine these spectra visually
for the presence of significant stellar features.
From this sample, we find 103 possible and 58
probable star--galaxy superpositions, indicating
a total CMASS star--galaxy
superposition rate of between 0.1\% and 0.2\%.
These star--galaxy superpositions that are given a spectroscopic
class of GALAXY are a source of sample impurity, as the galaxies
are typically neither bright enough nor of the correct color to fall
within the CMASS color--magnitude selection cuts on their own.
Any star--galaxy superpositions classified as STAR by the pipeline
are excluded from the large-scale structure analysis
and contribute only to target-selection inefficiency.

Our visual inspection set also contains 568 LOWZ galaxies brighter than the
fiber-magnitude threshold.  All of these spectra are confidently classified and
redshifted by both the pipeline and the visual inspection, with three
classified as stars.  This is consistent with the automated completeness
and stellar contamination rate for the full LOWZ sample, with
no detectable incidence of catastrophic failures.

\subsection{Galaxy redshift precision}

Redshift errors are calculated from the curvature of
the $\chi^2$ function in the vicinity of the minimum
value that is used to determine the best-fit redshift measurement.
To assess the accuracy of these statistical error estimates,
we make use of a set of 27170 repeat observations of CMASS targets
and 7503 repeat observations of LOWZ targets within the DR9 data set.
We reference all repeat observations
to the \texttt{SPECPRIMARY} observation of a given object,
and scale the redshift difference between the two observations
by the quadrature sum of the error estimates from the two epochs.
We then construct a histogram of these scaled velocity differences and fit it
with a Gaussian function.  If the estimated errors accounted
for all the statistical uncertainty, these fitted Gaussians would have
a dispersion parameter of unity.  Figure~\ref{fig:z_err_hist} shows the results
of this analysis, with fitted dispersions of $\sigma = 1.34$ for the
LOWZ sample and $\sigma = 1.19$ for the CMASS sample.  Thus, while
slightly underestimated, the redshift errors are impressively close to
being statistically accurate.  The greater scatter (relative
to the statistical error estimates) for the LOWZ sample
suggests that systematic effects become more important at higher
S/N\@.

This analysis of repeat spectra also yields
44 CMASS re-observations that have absolute
redshift differences $|\Delta z| > 0.005$ between the two epochs.
These are primarily due to galaxy--galaxy superpositions
at distinct redshifts, un-masked spectrum artifacts, and
a number of type II quasars for which broad [O\textsc{iii}]~5007 emission
is confused with H$\alpha$ in one epoch but not the
other (see Item~\ref{item:type2} in \S\ref{sec:issues}).
The implied 0.16\% CMASS
redshift impurity rate is consistent with the value found from
the truth-table tests of \S\ref{subsec:galcomp}.
For the LOWZ repeat observations, two spectra yield $|\Delta z| > 0.005$,
both of which are galaxy--galaxy superpositions.

For all CMASS and LOWZ targets, we also compute the distribution of
estimated redshift errors as a function of redshift.
These distributions are shown in Figure~\ref{fig:v_err_hist}.
In all cases, typical errors are a few tens of km\,s$^{-1}$ even when scaled
up to reflect the super-statistical scatter displayed in Figure~\ref{fig:z_err_hist}.
These errors are well below the 300\,km\,s$^{-1}$ redshift precision
requirement of the BOSS galaxy large-scale structure
science analyses.

\subsection{Galaxy redshift success dependence}
\label{subsec:zfaildepend}

\begin{figure}[t]
\epsscale{1.2}
\plotone{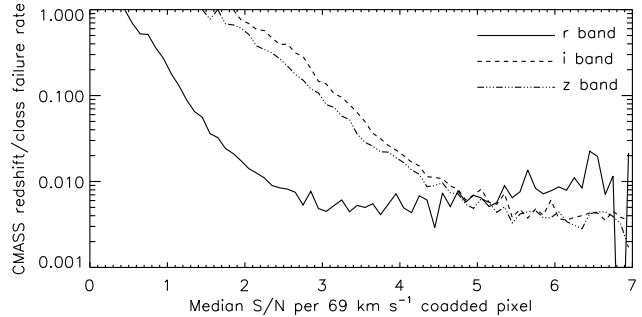}
\caption{\label{fig:sdss_snr_zrate}
BOSS CMASS sample redshift and classification failure rate
(i.e., \texttt{ZWARNING\_NOQSO > 0})
as a function of median spectroscopic S/N within the
SDSS $r$ (solid), $i$ (dashed), and $z$ (dot-dashed)
bandpass regions of the spectrum.}
\end{figure}

As in any redshift survey,
spectroscopic S/N is the primary determinant of
redshift success in BOSS\@.
Figure~\ref{fig:sdss_snr_zrate} shows the dependence of the
CMASS galaxy redshift failure rate as a function of the
median spectroscopic signal-to-noise ratio over the
SDSS $r$, $i$, and $z$ bandpass ranges, which represent the most relevant
regions of the spectrum for measuring continuum redshifts of passive
galaxies over the redshift interval $z \approx 0.4$--0.8.  Failure
is defined in the sense of \texttt{ZWARNING\_NOQSO > 0},
so that targets confidently identified as stars
are counted as a success for the pipeline even though they
represent a failure in the larger sense of galaxy targeting
and redshift measurement.  We see
a decrease in the failure rate as a function of $r$-band
S/N up to S/N$_r \simeq 3$, where an asymptotic
minimum of $\approx 5 \times 10^{-3}$ is reached.
For CMASS spectra with S/N$_r = 3$, the typical value for
both S/N$_i$ and S/N$_z$ is approximately 6, consistent
with the S/N values in those bands at which the
asymptotic failure rate is reached in Figure~\ref{fig:sdss_snr_zrate}.

\begin{figure}[t]
\epsscale{1.2}
\plotone{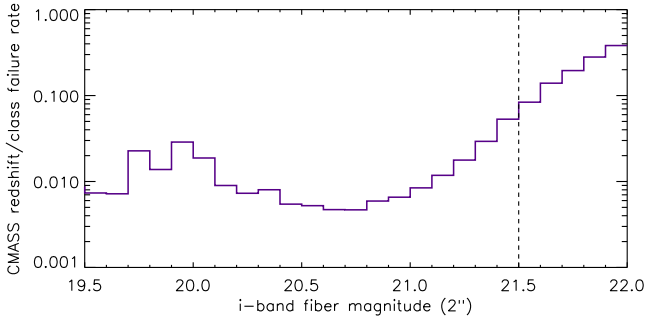}
\caption{\label{fig:ifibrate}
CMASS sample failure rate as a function of
apparent (not extinction-corrected)
$i$-band magnitude within the 2$^{\prime\prime}$-diameter
BOSS fiber, using the best single
spectroscopic observation of
each CMASS target in the DR9 data set.
Vertical dashed line at 21.5 indicates the nominal
fiber-magnitude faint limit of the CMASS sample.}
\end{figure}

Galaxy magnitude correlates strongly with
spectroscopic S/N and hence with redshift success:
this is the motivation for the formal CMASS sample
limit of $i$-band magnitude brighter than
21.5 within a 2$\arcsec$-diameter BOSS fiber.
To gauge the dependence of redshift
completeness on this limit,
Figure~\ref{fig:ifibrate} shows the
CMASS sample redshift failure rate as a function of
$i_{\mathrm{fiber}}$, selecting the best single
observation of each target.
Targets fainter than $i_{\mathrm{fiber}} = 21.5$ are
available from a more permissive CMASS cut
applied during commissioning observations.
At the formal CMASS cutoff, the marginal
failure rate is about 7\%.

\begin{figure}[b]
\epsscale{1.2}
\plotone{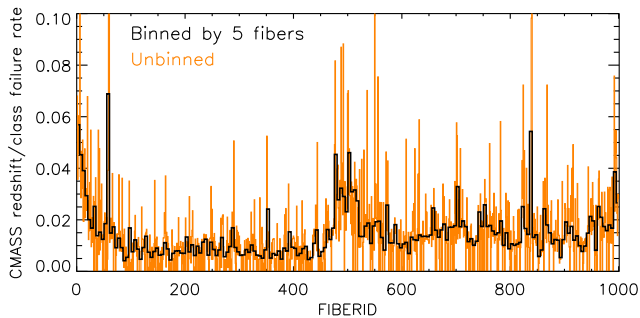}
\caption{\label{fig:fiberid}
CMASS sample failure rate as a function of fiber
number.  Generated for all spectroscopic
observations of CMASS sample targets with
$i_{\mathrm{fiber}} < 21.5$.  Large-scale structure
is due to spectrograph camera optics, and small-scale
peaks are associated with bad CCD columns.}
\end{figure}

The characteristics of the BOSS spectrograph
optics and CCD detectors produce a weak dependence of
redshift success rate on fiber identification
number along the linear
spectrograph slit-heads.
Figure~\ref{fig:fiberid}
presents this dependence for the CMASS sample.  This figure
is generated only for targets brighter than the
$i_{\mathrm{fiber}} < 21.5$ cut, but includes all survey
spectra of each target (i.e., no \texttt{SPECPRIMARY} cut)
so as to give an unbiased
picture of performance versus fiber number.
The upturns near fiber numbers 1, 500, and 1000 are
associated with the edges of the spectrograph
camera fields of view, and are described further
in Item~\ref{item:badfiber} in \S\ref{sec:issues} below.
The effects of isolated bad CCD columns are
also evident, and are described in Item~\ref{item:badcol}
in \S\ref{sec:issues}.  The failure rate is slightly
higher on average for fibers above 500, corresponding to
a lower end-to-end survey-averaged throughput for the optics and
CCDs of spectrograph 2 as compared to those of spectrograph 1.

\begin{figure}[t]
\epsscale{1.23}
\plotone{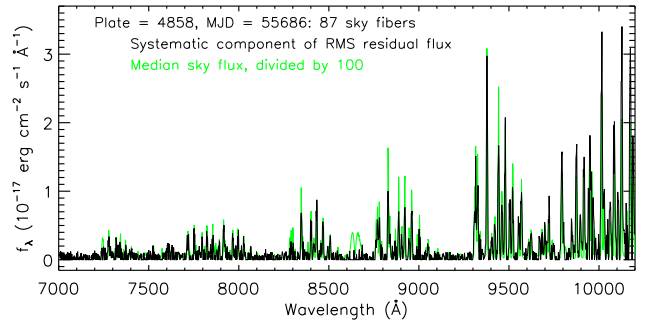}
\caption{\label{fig:skyres}
Systematic sky-subtraction RMS residual spectra (black line) computed
from sky-subtracted blank sky fibers of
a representative BOSS plate.
Estimated statistical errors have been subtracted in quadrature
from RMS residual flux at each wavelength.  Also shown is the
median sky flux spectrum scaled down by a factor of 100.}
\end{figure}

\begin{figure}[b]
\epsscale{1.2}
\plotone{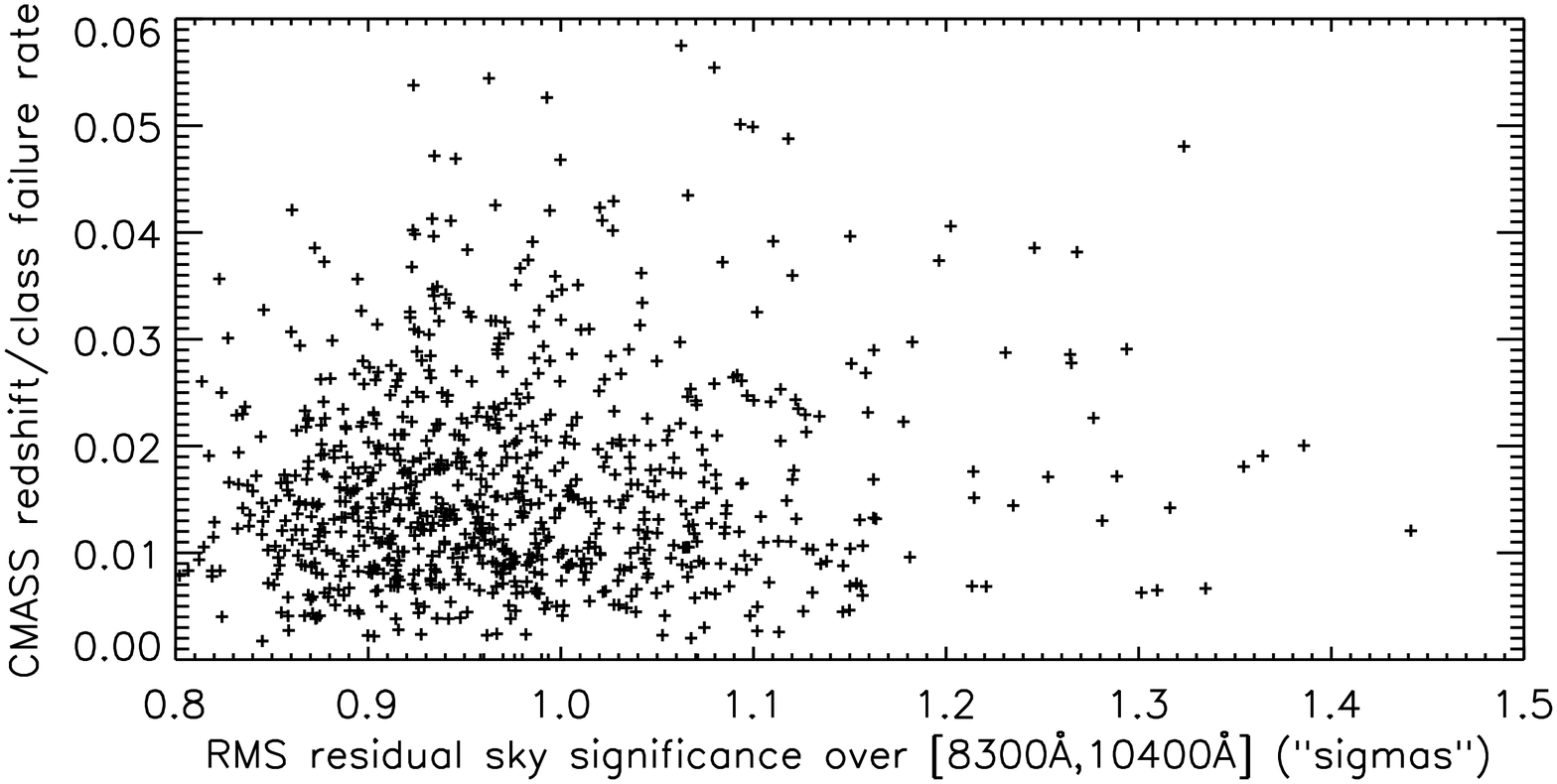}
\plotone{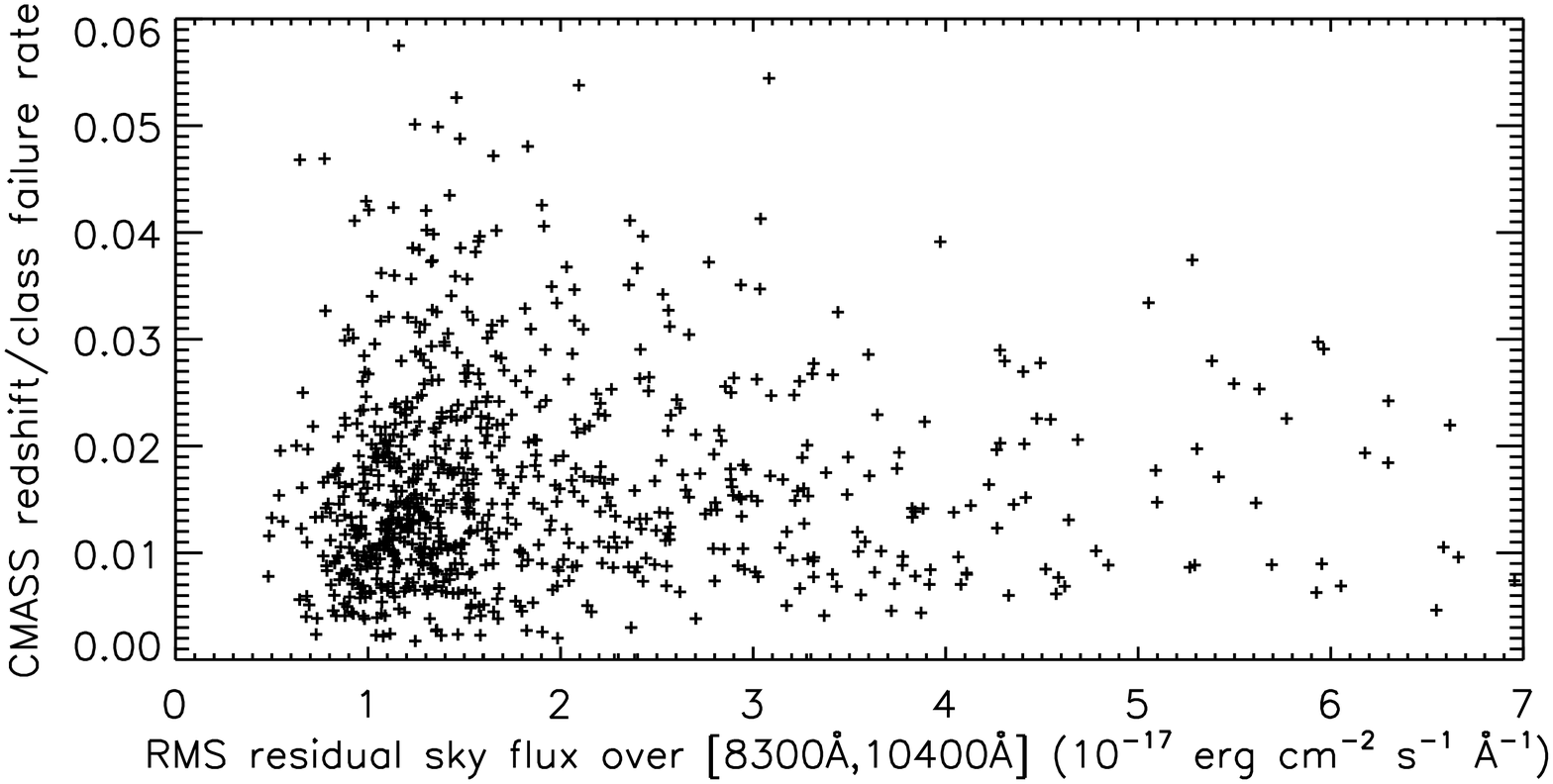}
\caption{\label{fig:sky_zrate}
CMASS sample redshift and classification failure rate
versus RMS residual flux in sky-subtracted sky fibers,
for BOSS plates with at least 300 CMASS galaxy sample
targets.  Each point represents one \texttt{PLATE-MJD}\@.
The top panel abscissa is in units of statistical
significance, while the bottom panel is in units of
specific flux.  No correlation is seen.
RMS significance values
of less than ``1-sigma'' reflect unaccounted
pixel-to-pixel correlations introduced by the re-binning and
co-addition of spectra.}
\end{figure}

In principle, variations in the quality of sky
foreground subtraction
can also affect spectroscopic redshift success.
In practice, we do not see this effect in BOSS\@.
Figure~\ref{fig:skyres} shows the spectrum of systematic
sky-subtraction residual flux measured from the sky-subtracted
blank-sky fibers of a representative BOSS plate, calculated
by subtracting statistical spectrum pixel error estimates
in quadrature from the root-mean-square (RMS) residual spectrum
across all sky fibers on the plate.  At the positions of
bright OH air-glow lines, the systematic residuals are generally
at or below 1\% of the sky flux.
To test whether the redshift failure rate is affected significantly
by variations in sky-subtraction quality, we quantify the level of residual
flux from the sky-subtraction process in each plate as the
RMS flux in all sky-subtracted blank-sky fibers over the
wavelength range 8300\AA\ to 10400\AA, where the
effects of OH air-glow lines are particularly pronounced.
Figure~\ref{fig:sky_zrate} displays the results of this
test, with RMS residual flux expressed both in units
of estimated statistical significance and in units of
specific flux.  In both cases, there is no discernible correlation between
sky-subtraction residual scale and redshift failure rate.
The two conclusions we draw are that (1) the quality of
BOSS sky subtraction is uniformly high, and (2) residual variations
in the quality of this sky subtraction do not significantly affect
redshift measurement for the passive, continuum-dominated CMASS galaxies.

\begin{figure*}[t]
\epsscale{1.15}
\plottwo{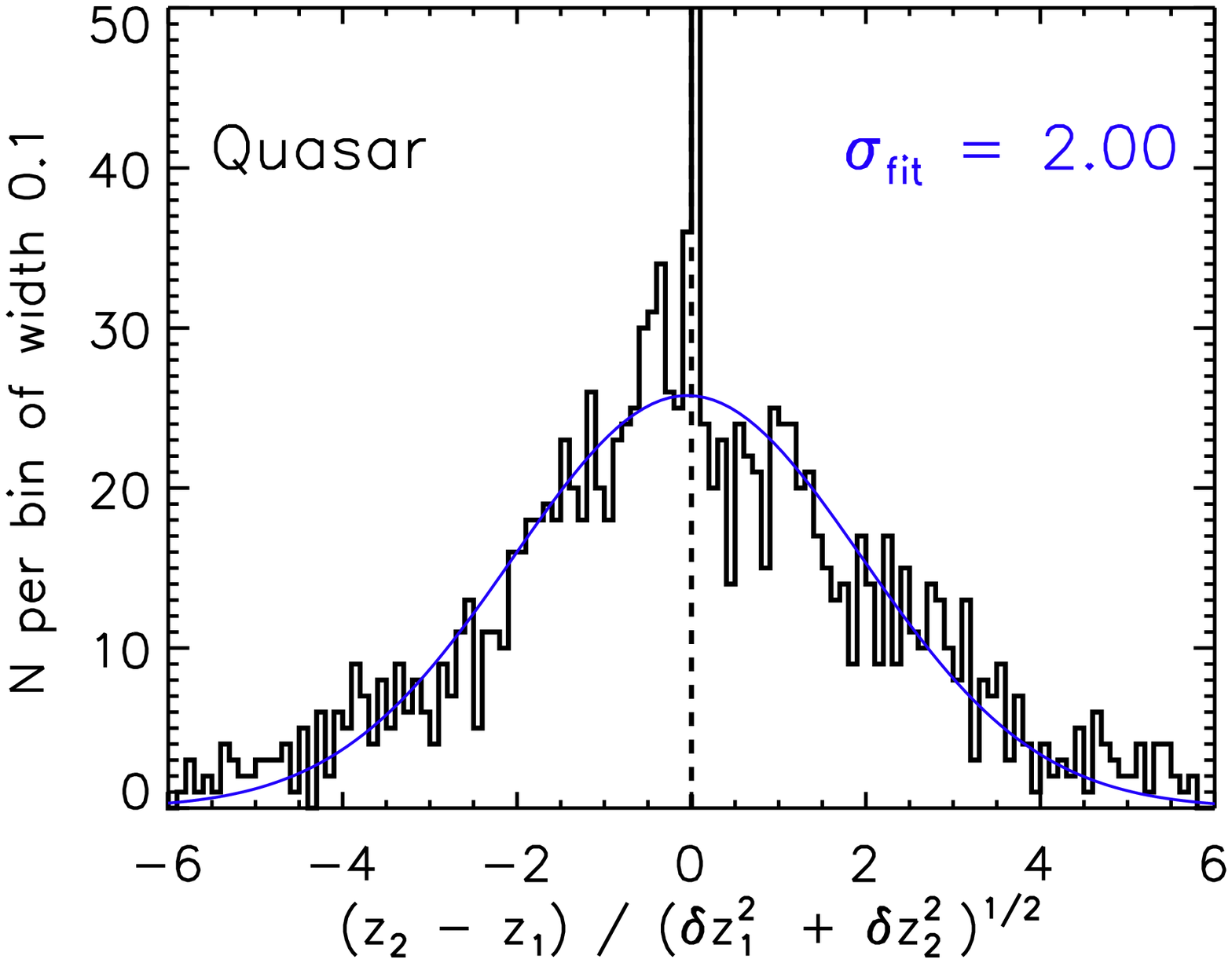}{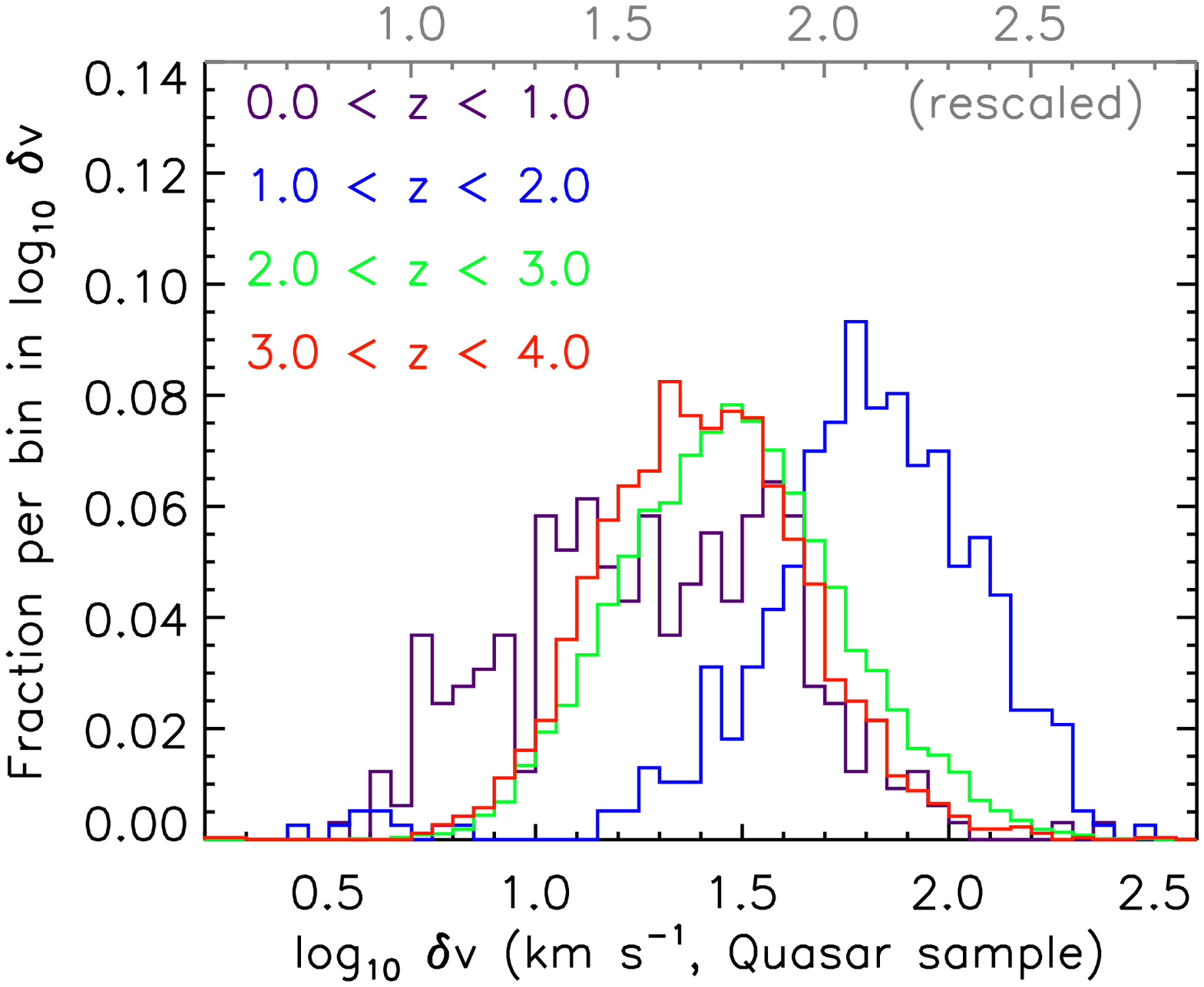}
\caption{\label{fig:z_err_hist_qso}
Distribution of error-scaled redshift differences between
repeat observations of BOSS quasars \textit{(left)},
and distribution of estimated single-epoch
\textit{statistical} quasar redshift errors
for multiple redshift ranges \textit{(right)}.
True quasar redshift errors are likely dominated
by systematic effects not reflected here.
The upper and lower horizontal axes in
the right-hand plot are as in
Figure~\ref{fig:v_err_hist}.
}
\end{figure*}

\subsection{Quasar redshift success}

\begin{figure}[b]
\epsscale{1.23}
\plotone{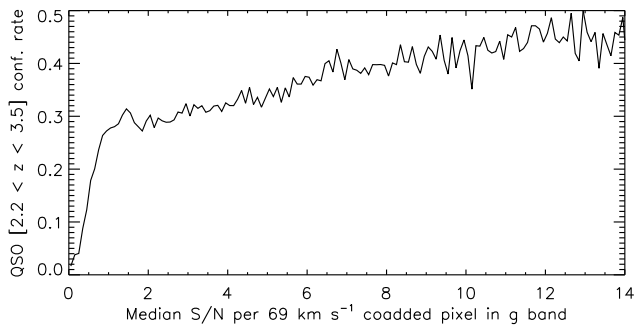}
\caption{\label{fig:qso_conf_rate}
Spectroscopic confirmation rate of quasars
with redshift $2.2 < z < 3.5$ from among
BOSS quasar sample targets, as a function of
median $g$-band S/N per spectroscopic pixel.}
\end{figure}

Unlike the BOSS galaxy samples,
the BOSS quasar sample does not have a stated requirement
on automated classification and redshift success.
The entire quasar target sample is being manually
inspected to provide a catalog
of visually verified classifications and redshifts
\citep{Paris12}, for which the automated BOSS
pipeline redshifts provide the initial default value.
From Table~\ref{table:dr9summary}, we find that the \texttt{idlspec2d}
pipeline reports a confident classification
and redshift (i.e., \texttt{ZWARNING == 0})
for about 79\% of the unique spectra
of the BOSS quasar target sample.  The majority of the remaining 21\% of the
quasar sample observations are low-S/N spectra of faint targets.
Approximately 51.5\% of the unique observed
quasar sample targets are spectroscopically confirmed as
quasars; most of the confidently classified
non-quasar spectra are stars (typically of spectral type F)
occupying the same region of color space as quasars in the
targeted redshift range.
However, only 33.6\% of the
unique target sample are confirmed quasars at the
redshifts $2.2 < z < 3.5$ which are the focus of the BOSS Ly$\alpha$
forest analysis \citep{Dawson12}.
Figure~\ref{fig:qso_conf_rate} presents the spectroscopic
confirmation rate for quasars in this redshift range
$2.2 < z < 3.5$, as a function of median S/N per pixel
over the $g$-band wavelength range.

The full comparison of
visual redshifts and pipeline redshifts
for BOSS quasar-sample targets
is presented in \citet{Paris12}, and is beyond the scope
of this current work.  We note two particular statistics here.
First, the visual inspections provide a
1.7\% increase in the sample of $2.2 < z < 3.5$ quasars beyond those
that are confidently identified by
the automated pipeline.  Second, 0.6\% of the quasars
identified confidently by the pipeline at $2.2 < z < 3.5$
either have redshifts in disagreement
by $|\Delta z| > 0.05$ with the visual-inspection values,
or do not have confident visual identification
despite having been inspected.
The latter are due mostly to extremely
broad absorption-line quasars and to line mis-identifications.
The overall conclusion, however, is that the completeness
and purity of the automated quasar classification and
redshift measurement is quite high.

Figure~\ref{fig:z_err_hist_qso} shows the distribution of
error-scaled redshift differences for 1464 repeat BOSS observations
of confirmed quasars, as well
as the redshift-dependent distributions of statistical single-epoch redshift
error estimates, analogous to
Figures~\ref{fig:z_err_hist} and \ref{fig:v_err_hist} for galaxies.
For quasars, the \textit{statistical}
pipeline redshift errors are underestimated by a factor
of approximately two, although the true
errors in the pipeline quasar redshifts are likely dominated
by systematic effects.  Eight of the repeat observations, or about 0.5\%,
give a redshift difference of $|\Delta z| > 0.05$, consistent
with the rate of catastrophic errors found by the comparison
with the visual inspections.
The redshift range 1.0--2.0 is particularly difficult since the
observed optical spectra
do not have either the narrow [O\textsc{iii}] 5007 line
or the strong Ly$\alpha$ line to guide the template fit.

\subsection{Stellar radial velocity precision}

\begin{figure}[t]
\epsscale{1.2}
\plotone{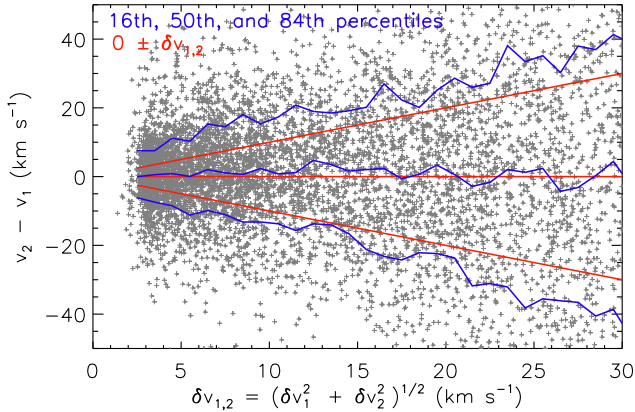}
\caption{\label{fig:starvdiff}
Velocity difference between epochs for 8174
stars with more than one good spectrum
in the BOSS data set, as a function of the quadrature sum of
statistical error estimates from the two differenced epochs.
Also plotted are the 16$^{\mathrm{th}}$, 50$^{\mathrm{th}}$, and 84$^{\mathrm{th}}$ percentile curves of this
velocity difference \textit{(blue)} and the expected
statistical $0 \pm 1 \sigma$ lines \textit{(red)}.}
\end{figure}

We now briefly examine the precision and accuracy of BOSS stellar radial velocities
based on stellar repeat observations.  Specifically, we identify 8174
repeat observations of objects classified as \texttt{STAR}
with \texttt{ZWARNING}~$== 0$ for both epochs.
In Figure~\ref{fig:starvdiff}, we plot the velocity difference
between the two epochs of these repeats against the quadrature
sum of their statistical error estimates.  We see that the
distribution becomes tighter at higher S/N as
expected, with reasonably good agreement between estimated
statistical error and actual velocity differences above
approximately 15\,km\,s$^{-1}$ in combined statistical error
(or approximately 10\,km\,s$^{-1}$ in single-epoch error).
Subtracting the statistical error estimates in quadrature
from the half-difference between the 84$^{\mathrm{th}}$ and 16$^{\mathrm{th}}$
percentile velocity differences, and dividing by a
factor of $\sqrt{2}$ to convert to a single-epoch value,
we find a systematic radial-velocity floor of
approximately 4.5\,km\,s$^{-1}$ at the high S/N end,
comparable to the 4\,km\,s$^{-1}$ precision attained for bright
stars by the SEGUE project in SDSS-I/II \citep{Yanny09}.

\section{Known Issues}
\label{sec:issues}

\begin{figure*}[t]
\epsscale{1.1}
\plotone{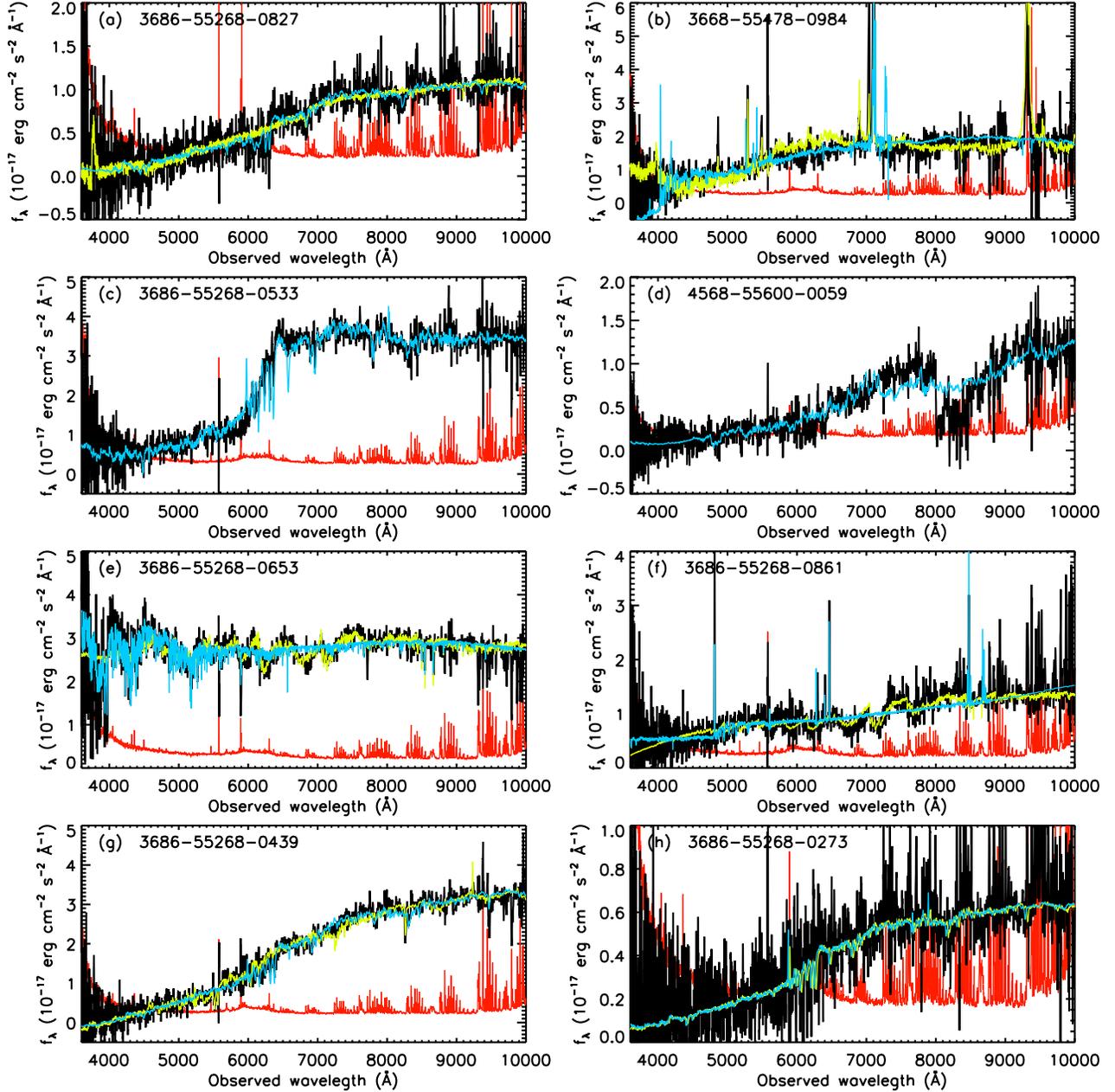}
\caption{\label{fig:crappo} 
Mosaic of problematic BOSS spectra.  Black lines show data (smoothed
over a 5-pixel window), and red lines show 1-$\sigma$ noise level
estimated by the extraction pipeline.
Spectra are labeled by \texttt{PLATE-MJD-FIBERID}.
All spectra are from the CMASS sample.  Individual objects are:
(a) redshift $z = 0.589$ galaxy
for which the overall minimum-$\chi^2$ fit is an
unphysical quasar-class model (yellow), but for which the \texttt{NOQSO}
redshift and class (cyan) are confident and correct, as described
in \S\ref{subsec:z_noqso}; (b) type-II quasar
with a correct quasar-class redshift of $z = 0.419$ (yellow) but an incorrect
\texttt{NOQSO} redshift $z = 0.083$ (cyan) due to confusion of
broad [O\textsc{iii}]~5007 with H$\alpha$; (c) spectrum with an
exaggerated break feature at the 6000\,\AA\ dichroic transition, due to cross-talk
effects from a bright star in a neighboring fiber, but for which the
pipeline redshift of $z = 0.603$ (cyan) is nevertheless correct;
(d) Spectrum affected by a transient bad CCD column in the
region 8000\,\AA\,$<\lambda<$\,9000\AA, with an unphysical
galaxy-class model (cyan) for which the \texttt{SMALL\_DELTA\_CHI2}
bit is set in the \texttt{ZWARNING\_NOQSO} mask; (e) spectral
superposition of a G star with an M star (the spectrum is
confidently classified
as \texttt{STAR}), with the best-fit
G-star-plus-polynomial shown in cyan and the best-fit
M-star-plus-polynomial shown in yellow;
(f) spectral superposition between a redshift $z = 0.291$
emission-line galaxy (cyan) and an M star (yellow);
(g) spectral superposition between a redshift
$z = 0.606$ absorption-line galaxy (cyan) and a redshift
$z = 0.402$ absorption-line galaxy (yellow);
(h) spectrum of a galaxy for which the pipeline cannot
distinguish with statistical confidence between a
redshift of $z = 0.576$ (cyan) and $z = 0.582$ (yellow,
largely hidden by cyan), and for
which the \texttt{SMALL\_DELTA\_CHI2}
bit is consequently set in the
\texttt{ZWARNING\_NOQSO} mask (see \S\ref{subsec:zmeasure}
and \S\ref{subsec:z_noqso}) since
these two redshifts differ by more than 1000\,km\,s$^{-1}$.}
\end{figure*}

In order to freeze a set of reductions for collaboration
analysis and public release, we have accepted the presence
of a number of known outstanding
issues in the software that either were deemed small enough in a statistical
sense within the survey, or were discovered after the software
freeze deadline.  These issues are documented in the following list,
and several are illustrated in Figure~\ref{fig:crappo}.

\begin{enumerate}

\item PCA fits of the \texttt{GALAXY} and \texttt{QSO}
classes can sometimes yield
unphysical basis combinations at low S/N\@.  This effect is part of the
motivation for the \texttt{Z\_NOQSO} redshifts described
in \S\ref{subsec:z_noqso}, and is illustrated in
panel ``a'' of Figure~\ref{fig:crappo}.  In order to enable a targeting-blind
spectroscopic classification of the sort used in SDSS-I/II,
this effect could be remedied by
priors on physical PCA coefficient combinations, or by
non-negativity requirements on archetype-based models
such as are used for non-CV stellar classifications
in \texttt{idlspec2d}.
These alternatives are the subject of ongoing development for future BOSS
data releases.

\item \label{item:type2} A small number of type II quasars
\citep[e.g.,][]{Zakamska03}
at redshift $z \sim 0.5$ are selected by the CMASS cuts
due to their colors, but their obscured-AGN spectra are not
typical of the majority of galaxies used to train the
galaxy redshift templates.  The inclusion of several such systems
in the galaxy-template training set has addressed this issue
partially, but a number of these objects have a best-fit galaxy-template
redshift that confuses broad [O\textsc{iii}]~5007 for H$\alpha$.
Their quasar-template redshifts are generally correct, but due to
the \texttt{Z\_NOQSO} redshift strategy employed for the BOSS
galaxy samples (\S\ref{subsec:z_noqso}), their adopted redshifts are often in error
(see panel ``b'' of Figure~\ref{fig:crappo}.)
Since these objects represent such a small percentage of the
BOSS galaxy target samples, these errors were deemed acceptable
for DR9 galaxy-clustering analyses.

The fundamental problem is that the spectra of type II quasars are
sufficiently different from the spectra of most BOSS galaxies that
we cannot span the space of both categories
with the current number of PCA
templates (four) in the single \texttt{GALAXY} basis set.
In future BOSS data releases, we anticipate addressing
this issue through either higher-dimensional basis sets
with physical coefficient priors, sub-division of the \texttt{GALAXY}
class into several subclasses each with its own
basis set, or an archetype-based galaxy redshifting algorithm.

\item \label{item:crosstalk} A small number of spectra are affected
by cross-talk from bright stars (generally
spectrophotometric standards) in neighboring fibers.
This is often manifested in a strong break feature at the
dichroic transition around 6000\,\AA\
(see panel ``c'' of Figure~\ref{fig:crappo}), due to
different levels of cross-talk between the red and blue
arms of the spectrograph \citep{Smee12}.
These effects appear to occur less frequently at later survey
dates, presumably because of improvements in
the operating focus of the BOSS spectrographs.
We intend to address these effects in future BOSS data releases
through improvements in the extraction codes,
and to flag any spectra that remain compromised.
No masking of this effect is implemented for
BOSS DR9 data, however, except to the extent
that it sometimes triggers a \texttt{ZWARNING} flag.

\item \label{item:badfiber} As discussed in
\S\ref{subsec:zfaildepend} and shown
in Figure~\ref{fig:fiberid}, the BOSS redshift
success rates are somewhat dependent on fiber number
in the sense that fibers near the edge of the spectrograph camera fields
of view (\texttt{FIBERID} values near 1, 500, and 1000)
have lower success rates.
Longer-term development of new extraction codes based on the 2D PSF-modeling
approach of \citet{Bolton10} is ongoing, and may mitigate this problem to
a significant extent.

\item \label{item:badcol} A few columns in the BOSS CCDs are bad only in a transient
sense, and are not included in the bad-column masks applied
to the CCD frames.  These columns lead to occasional spectrum artifacts
concentrated near particular fiber numbers (see panel ``d'' of Figure~\ref{fig:crappo})
that are not masked or flagged.

\item White-dwarf, L-dwarf, carbon-star, and cataclysmic-variable star subclasses have
less accurate template radial-velocity zero-points in comparison to
the stellar archetypes derived from the Indo-U.S\@. library.  This issue may
be rectified in future data releases, although
the primary role of stellar templates in BOSS will
remain to correctly classify
and set aside non-galaxies and non-quasars.

\item Spectra showing superpositions of two objects are not systematically
identified and flagged by the pipeline.  While the majority of
BOSS spectra are of single objects, superpositions are occasionally
found to occur.  In some cases, the inclusion of the polynomial terms
in the redshift model fitting leads to fits of almost equal quality
for the two components individually, leading to a \texttt{SMALL\_DELTA\_CHI2} flag
in the \texttt{ZWARNING} (or \texttt{ZWARNING\_NOQSO}) mask.
In other cases, one component is dominant and is identified by
the pipeline as the confident classification and redshift,
but with the second component typically identified by
one of the lower-quality fits reported in the \texttt{spZall} file.
Various examples of superposition spectra are displayed in Figure~\ref{fig:crappo},
including star--star (panel ``e''), star--galaxy (panel ``f''),
and galaxy--galaxy (panel ``g'').  A systematic search for superposition
spectra in the BOSS data set by the BOSS Emission-Line Lens
Survey (BELLS, \citealt{Brownstein12}) has discovered a large
sample of strong gravitational lens galaxies.

\end{enumerate}

\section{Summary and Conclusion}
\label{sec:summary}

We have described the ``1D'' component of the
\texttt{idlspec2d} pipeline that provides automated
redshift measurement and and classification for the SDSS-III BOSS DR9
data set, which comprises 831,000 optical spectra.
This software is substantially similar to the \texttt{idlspec2d}
redshift analysis code used for SDSS-I/II data, but has been upgraded with new
templates and several new algorithms for application to the BOSS project,
and has been presented in great detail for the first time in this work.
The pipeline also provides additional parameter measurements, including
emission-line fits for all objects,
and velocity-dispersion likelihood curves for objects classified as galaxies.
The redshift success rate
of the \texttt{idlspec2d} pipeline is well in excess of the
scientific requirements of the BOSS project.  The software provides
first-principles estimates of statistical
redshift errors that are Gaussian distributed and accurate to within
small correction factors.  The ``2D'' component of the \texttt{idlspec2d}
pipeline that extracts spectra from raw CCD pixels is the subject
of \citet{Schlegel12}.
Full data-model information for both the 2D and 1D BOSS pipeline outputs
can be found at the SDSS-III DR9 website (\url{http://www.sdss3.org/dr9/}).

Development work continues on data-reduction software for BOSS,
both in the calibration and extraction of spectra, and in the classification
and redshift analysis procedures.  Subsequent BOSS data releases will be accompanied
by similar documentation of the implemented results of this ongoing development.

\acknowledgments

Funding for SDSS-III has been provided by the Alfred P. Sloan Foundation, the Participating Institutions, the National Science Foundation, and the U.S. Department of Energy Office of Science. The SDSS-III web site is \url{http://www.sdss3.org/}.

SDSS-III is managed by the Astrophysical Research Consortium for the Participating Institutions of the SDSS-III Collaboration including the University of Arizona, the Brazilian Participation Group, Brookhaven National Laboratory, University of Cambridge, Carnegie Mellon University, University of Florida, the French Participation Group, the German Participation Group, Harvard University, the Instituto de Astrofisica de Canarias, the Michigan State/Notre Dame/JINA Participation Group, Johns Hopkins University, Lawrence Berkeley National Laboratory, Max Planck Institute for Astrophysics, Max Planck Institute for Extraterrestrial Physics, New Mexico State University, New York University, Ohio State University, Pennsylvania State University, University of Portsmouth, Princeton University, the Spanish Participation Group, University of Tokyo, The University of Utah, Vanderbilt University, University of Virginia, University of Washington, and Yale University.

This research has made use of the POLLUX database
(\texttt{http://pollux.graal.univ-montp2.fr})
operated at LUPM  (Universit\'{e} Montpellier II - CNRS, France)
with the support of the PNPS and INSU\@.



\end{document}